\pdfoutput=1
\documentclass[preprint,pre,floats,showkeys,aps,amsmath,amssymb,superscriptaddress,11pt,nofootinbib]{revtex4-1}
\usepackage[utf8]{inputenc}
\usepackage{graphicx}
\usepackage{natbib}
\setcitestyle{square,comma,numbers,sort&compress}
\usepackage{amsmath,amsfonts,amssymb}
\usepackage{array}
\usepackage{tabularx}
\usepackage{enumerate}
\usepackage[left = 1.5in, top=1.5in,right=1.5in,bottom=1.5in,a4paper]{geometry}
\usepackage[dvipsnames]{xcolor}
\usepackage{colortbl}
\usepackage{comment}
\usepackage{siunitx}
\usepackage{lineno}
\usepackage{tikz}
\usepackage[compat=1.1.0]{tikz-feynman}
\usepackage{cancel}
\usepackage{hyperref}

\newcommand{\be}{\begin{equation}}
\newcommand{\ee}{\end{equation}}
\newcommand{\ben}{\begin{equation*}}
\newcommand{\een}{\end{equation*}}

\def\bea{\begin{eqnarray}}
\def\eea{\end{eqnarray}}
\def\bean{\begin{eqnarray*}}
\def\eean{\end{eqnarray*}}

\def\l2{\log_2\,}
\newcommand{\barr}{\begin{array}}
\newcommand{\earr}{\end{array}}

\newcommand{\bed}{\begin{displaymath}}
\newcommand{\eed}{\end{displaymath}}
\newcommand{\bal}{\begin{array}{ll}}
\newcommand{\eal}{\end{array}}

\def\sc#1{\textstyle{\scriptstyle #1}}

\def\ket#1{\vert\,#1\,\rangle}

\def\mc#1{\mathcal#1}

\newcolumntype{Y}{>{\centering\arraybackslash}X}
\newcommand{\g}[1]{\mathbf{#1}}
\newcommand{\vev}[1]{\langle #1 \rangle_0}
\newcommand{\gb}[1]{\bar{\mathbf{#1}}}
\newcommand{\Higgs}{\bar{H}_{\g{5}}}

\newcommand{\vphi}{\varphi^{(t)}_{\g{3}_1}}

\newcommand{\matc}[1]{  \left(
\begin{array}{c}
 #1
\end{array}
\right)
}

\newcommand{\blu}[1]{\textcolor{black}{#1}}
\newcommand{\myred}[1]{\textcolor{black}{#1}}
\newcommand{\myboxed}[1]{%
  \rlap{\hspace*{\dimexpr\fboxrule+\fboxsep\relax}%
    \phantom{\m@th$\displaystyle#1$}}%
    \smash{\boxed{#1}}}

\renewcommand{\thesection}{\arabic{section}{\Large}}
\renewcommand{\thesubsection}{\thesection.\arabic{subsection}}
\renewcommand{\thesubsubsection}{\thesubsection.\arabic{subsubsection}}

\makeatletter
\renewcommand{\p@subsection}{}
\renewcommand{\p@subsubsection}{}
\makeatother
\begin{document}
%\today \hspace{7.5cm} \textcolor{red}{version: 5.3}

\title{\Large Tribimaximal Mixing in the $SU(5) \times \mc T_{13}$ Texture \\\vspace*{1cm}}
%\author{\small{M. Jay P\'erez, Moinul Hossain Rahat, Pierre Ramond, Alexander Stuart, Bin Xu}}

\author{M. Jay P\'erez} 
\email[Email: ]{mperez75@valenciacollege.edu}
\affiliation{\small{Valencia College, Osceola Science Department, Kissimmee, FL 34744, USA}}

\author{Moinul Hossain Rahat} 
\email[Email: ]{mrahat@ufl.edu}
\affiliation{\small{Institute for Fundamental Theory, Department of Physics,
University of Florida, Gainesville, FL 32611, USA }}

\author{Pierre Ramond} 
\email[Email: ]{ramond@phys.ufl.edu}
\affiliation{\small{Institute for Fundamental Theory, Department of Physics,
University of Florida, Gainesville, FL 32611, USA }}

\author{Alexander J. Stuart} 
\email[Email: ]{astuart@ucol.mx}
\affiliation{\small{Facultad de Ciencias-CUICBAS, Universidad de Colima, C.P. 28045, Colima, M\'exico}\\\vspace*{1cm}}
%\affiliation{\small{Dual CP Institute of High Energy Physics, C.P. 28045, Colima, Mexico}\\\vspace*{1cm}}

\author{Bin Xu } 
\email[Email: ]{binxu@ufl.edu}
\affiliation{\small{Institute for Fundamental Theory, Department of Physics,
University of Florida, Gainesville, FL 32611, USA }}

%\vskip .2cm
%\centerline{\today}
%\vskip 1cm

%\begin{flushright}
%{\jobname}
%\framebox{~~ \ampmtime~--~ \today { }}
%\end{flushright}

\begin{abstract}
\vskip 0.5cm
\noindent \blu{We extend the recently proposed $SU(5) \times \mc T_{13}$ model for the asymmetric texture to the up-type quark and seesaw sectors. The hierarchical up-type quark masses are generated from higher-dimensional operators involving family-singlet Higgses, gauge-singlet familons, and vector-like messengers. The complex-tribimaximal (TBM) seesaw mixing arises from the vacuum structure of a minimal number of familons, resulting in an alignment between the Yukawa and Majorana matrices of the seesaw formula. Introducing four right-handed neutrinos, normal ordering of the light neutrino masses is obtained, with $m_{\nu_1} = 27.6\ \mathrm{meV}$, $m_{\nu_2} = 28.9\ \mathrm{meV}$ and $m_{\nu_3} = 57.8\ \mathrm{meV}$. Their sum almost saturates Planck's cosmological upper bound ($120$ $\text{meV}$). The right-handed neutrino masses are expressed in terms of two parameters for a particular choice of familon vacuum alignment. We predict the $\cancel{CP}$ Jarlskog-Greenberg invariant to be $|\mc J| = 0.028$, consistent with the current PDG estimate, and Majorana invariants $|\mc I_1| = 0.106$ and $|\mc I_2| = 0.011$. A sign ambiguity in the model parameters leads to two possibilities for the invariant mass parameter $|m_{\beta \beta}|$: $13.02$ or $25.21$ $\text{meV}$, both within an order of magnitude of the most rigorous experimental upper limit ($61$--$165$ $\text{meV}$).} 
\end{abstract}

\maketitle
\section{Introduction}
\blu{In Ref.~\cite{rrx2018asymmetric}, three of us proposed a minimally asymmetric Yukawa texture for the down-type quark matrix, $Y^{(-\frac{1}{3})}$, and charged lepton matrix, $Y^{(-1)}$, in the context of $SU(5)$ gauge-unification. Assuming a diagonal up-type quark Yukawa matrix $Y^{(\frac{2}{3})}$, this texture successfully reproduces the quark mixing angles and the mass ratios of the down-type quarks and charged leptons in the deep ultraviolet. The PMNS lepton mixing matrix bridges the $\Delta I_w = \frac{1}{2}$ physics of charged leptons to the unknown $\Delta I_w = 0$ physics of the seesaw sector:
\begin{align}
    \mc U_{PMNS} = \mc U^{(-1)^\dagger}\ \mc U_{seesaw}.
\end{align}
The large atmospheric and solar angles in the PMNS matrix are explained by Tribimaximal (TBM) \cite{harrison2002tri, *tbm2, *xing2002nearly, *he2003some, *wolfenstein1978oscillations} seesaw mixing, whereas the small reactor angle emerges entirely from the ``Cabibbo haze" \cite{cabibbohaze, *everett2006viewing, *Everett:2006fq, *kile2014majorana} provided by the charged leptons. Adding a single $\cancel{CP}$ phase \cite{Everett:2019idp,*parida2019high, *Shimizu2019,  *ballett2014testing, *ballett2014testingatm, *antusch2018predicting, *king2019theory, *ahriche2018mono, *delgadillo2018predictions, *girardi2016leptonic, *ding2019status, *Chen:2019egu,  *liu2019further, *petcov2018discrete, *girardi2015predictions, *girardi2015determining, *girardi2016predictions, *dinh2017revisiting, *penedo2018low, *agarwalla2018addressing, *petcov2016theory, *ge2011z2, *ge2012residual, *Ding:2019zhn, *CarcamoHernandez:2018djj, *Chen:2015siy, *Chen:2018eou} to TBM reproduces all three angles within $1\sigma$ of their PDG global fits \cite{tanabashi2018review}. Moreover, the phase yields the $\cancel{CP}$ Jarlskog-Greenberg invariant \cite{jarlskog1, *greenberg1} to be $|\mc J| = 0.028$, consistent with PDG \cite{tanabashi2018review}. }

%The nontrivial properties of the texture, e.g., the asymmetric term, equality of the determinants of the Yukawa matrices of the down-type quarks and charged-leptons, and separation of the $SU(5)$- $\gb{5}$ and $\overline{\g{45}}$ couplings,  were explained with an underlying $\mc T_{13}$ family symmetry in Ref.~\cite{Perez:2019aqq}. What remained unexplained were the origin of the TBM-seesaw mixing and the particular hierarchical structure of the up-type quark Yukawa texture. 

In Ref.~\cite{Perez:2019aqq}, we introduced a model where the ``fine-tunings'' of the asymmetric texture are upgraded to ``natural'' relations with the addition of a discrete family symmetry (see \cite{king2013neutrino, *tanimoto2015neutrinos, *meloni2017gut, *petcov2017discrete} and the references therein) $\mc T_{13} = \mc Z_{13} \rtimes \mc Z_{3}$, the smallest subgroup of $SU(3)$ with two inequivalent triplets \cite{miller1961theory, *fairbairn1964finite, *ludl2011comments,  *grimus2014characterization, *ishimori2010non, *kajiyama2011t13}, which are necessary to generate the asymmetry. Folded in with grand-unified theory (GUT) $SU(5)$, this model explains the features of the $\Delta I_w = \frac{1}{2}$ down-type quark and charged-lepton Yukawa matrices constructed from higher-dimensional operators in terms of gauge-singlet familons, family-singlet Higgses, and messengers with heavy vectorlike masses. \myred{A key feature of the model is ``crystallographic'' familon vacuum alignments, implying that all nonzero components of the triplet/antitriplet familons obtain the same order of vacuum expectation values.}

This paper expands the analysis to the up-type quark and seesaw sectors of the model. The up-type quark masses are explained by dimension-five, -six, and -seven operators, which yield a diagonal $Y^{(\frac{2}{3})}$ and reproduce their ultraviolet hierarchy.

Turning to the $\Delta I_w = 0$ seesaw sector, we show how the complex-TBM seesaw mixing arises from the vacuum structure of a minimal number of familons, resulting from the $\mc T_{13}$ Clebsch-Gordan coefficients.\footnote{Ref.~\cite{parattu2011tribimaximal} scans over subgroups of $SU(3)$ and identifies $\mc T_{13}$ as one of the groups that can yield TBM mixing. See \cite{ding2011tri, *hartmann2011neutrino, *PhysRevD.85.013012} for other approaches to study neutrino mixing with TBM in relation to $\mc T_{13}$ family symmetry as well as \cite{Feruglio:2019ktm} for a recent review of neutrino flavor symmetries.} It requires an alignment between the Yukawa ($Y^{(0)}$) and Majorana ($\mc M$) matrices of the seesaw formula
\begin{align}
    \mc S = Y^{(0)}\ \mc M^{-1}\ Y^{(0)^T},
\end{align}
\myred{without the need to specify familon vacuum expectation values}. The minimal construction with three right-handed neutrinos with TBM mixing yields mass relations between the light neutrinos incompatible with the oscillation data \cite{tanabashi2018review}. The addition of a gauge-singlet fourth right-handed neutrino is shown to produce TBM seesaw mixing, and $m_{\nu_2} = \frac{1}{2} m_{\nu_3}$ in two different scenarios, and by using the oscillation data \cite{tanabashi2018review} generates the three light neutrino masses in normal ordering: $m_{\nu_1} = 27.6$ $\text{meV}$, $m_{\nu_2} = 28.9$ $\text{meV}$ and $m_{\nu_3} = 57.8$ $\text{meV}$, with their sum close to the Planck value ($120$ $\text{meV}$) \cite{aghanim2018planck}.

%This construction predicts normal ordering of the light-neutrino masses: $m_{\nu_2} = \frac{1}{2} m_{\nu_3}$, but $m_{\nu_1}$ appears to be too large to be consistent with neutrino oscillation data \cite{tanabashi2018review}. }

%\blu{A minimal extension by introducing a fourth right-handed neutrino corrects $m_{\nu_1}$, while keeping the prized relationship between $m_{\nu_2}$ and $m_{\nu_3}$ intact. We discuss two scenarios where this is realized and predict $m_{\nu_1} = 27.6$ $\text{meV}$, $m_{\nu_2} = 28.9$ $\text{meV}$ and $m_{\nu_3} = 57.8$ $\text{meV}$. Sum of the three light-neutrinos (114.3$\text{meV}$) is very close to the cosmological upper bound (120$\text{meV}$) recently reported by the Planck collaboration \cite{aghanim2018planck}.} 

%\blu{Assuming simple vacuum alignment of the familons, we then calculate the right-handed neutrino masses in terms of two parameters. Our results show that these masses can be degenerate for a range of values of the parameters. }

The four right-handed neutrino masses are calculated in terms of two parameters assuming simple vacuum alignments of the seesaw familons. We find curious cases of degeneracies in their mass spectrum. 

We also calculate the $\cancel{CP}$ Dirac and Majorana phases \cite{schechter1982neutrinoless} yielded by the asymmetric texture with complex-TBM seesaw mixing. Together with the light neutrino masses, they predict the invariant mass parameter $|m_{\beta\beta}|$ in neutrinoless double-beta decay \cite{dell2016neutrinoless, *engel2017status, *vergados2016neutrinoless, *pas2015neutrinoless, *king2013power} to be either $13.02$ $\text{meV}$ or $25.21$ $\text{meV}$, \myred{depending on the sign of the parameters}, within an order of magnitude of the recently measured upper limit of $61$--$165$ $\text{meV}$ by the KamLAND-Zen experiment \cite{gando2016search}. 

The $SU(5) \times \mc T_{13}$ symmetry still allows for some unwanted tree-level vertices which can be prohibited by introducing a $\mc Z_n$ symmetry, where $n = 14$ or $12$ depending on which of the two aforementioned scenarios is realized in the seesaw sector. The full symmetry of the unified model is therefore $SU(5) \times \mc T_{13} \times \mc Z_n$, successfully explaining the masses and mixings of both quarks and leptons.

The organization of the paper is as follows. In Section \ref{sec:2}, we review the construction of the asymmetric texture, its key features, and how they are realized by a $\mc T_{13}$ family symmetry. Section \ref{sec:5} explains how the hierarchical up-type quark Yukawa texture is built from higher-dimensional operators. Section \ref{sec:3} discusses the seesaw sector in detail. In Section \ref{sec:4}, we calculate the Majorana phases and the invariant mass parameter $|m_{\beta\beta}|$. We summarize the unified model in Section \ref{sec:6}. Section \ref{sec:7} discusses the theoretical outlook and we conclude in Section \ref{sec:8}.

%%%%%%%%%%%%%%%%%%%%%%%%%%%%%%%%%%%%%%%%%%%%%%%%%%%%%%%%%%%%%%%%%%%%%%%%%%%%%%%%%%%%%%%%%%%%%%%%%%%%%%%%%

%%%%%%%%%%%%%%%%%%%%%%%%%%%%%%%%%%%%%%%%%%%%%%%%%%%%%%%%%%%%%%%%%%%%%%%%%%%%%%%%%%

\section{Asymmetric Tribimaximal Texture from $\mc T_{13}$}\label{sec:2}
\blu{In this section we review the key features of the asymmetric texture  and how it emerges from the discrete family symmetry $\mc T_{13}$.}  Our approach is inspired by ``gauge simplicity'' and ``seesaw simplicity'' in the deep ultraviolet. \blu{Gauge simplicity leads to $SU(5)$ grand unification of the Standard Model gauge groups, and relates $Y^{(-\frac{1}{3})}$ to $Y^{(-1)}$. Renormalization group running to the deep ultraviolet hints at suggestive relations between quark and charged-lepton masses: }
\begin{align}
    \frac{m_u}{m_c} \approx \frac{m_c}{m_t} \approx \lambda^4,\ \ \frac{m_s}{m_b} \approx \frac{\lambda^2}{3},\ \ \frac{m_e}{m_\tau} \approx \frac{\lambda^4}{9},\ \frac{m_\mu}{m_\tau} \approx \lambda^2,\ m_b \approx m_\tau, \label{GUTmass}
\end{align}
relating quark mass ratios to mixing angles through the Gatto relation \cite{gatto}
\begin{align}
    \sqrt{\frac{m_d}{m_s}} &\approx \lambda, \label{gattoeq}
\end{align}
and implies that
\begin{align}
    \det Y^{(-\frac{1}{3})} \approx \det Y^{(-1)}, \label{detcond}
\end{align}
where $\lambda \approx 0.225$ is the Wolfenstein parameter.
The GUT-scale relations of Eq.~\eqref{GUTmass} imply  $(m_\mu / m_s)( m_d / m_e) \approx 9$, which is in tension with the recent estimate  $(m_\mu / m_s)( m_d / m_e) \approx 10.7^{+ 1.8}_{-0.8}$ \cite{Antusch:2013jca}.

%We assume that these $SU(5)$-inspired input conditions are realized at the GUT scale ($\sim 10^{16}$ GeV); however, we do not elaborate how to approach the ultraviolet from the weak scale. For example, the success of $b$--$\tau$ unification at the GUT scale relies on the appearance of new colored states which couple differently to different flavors below $10^7$ GeV. We do not specify what those states are. We also do not assume supersymmetry in the form of MSSM, in which case the ratio $(m_\mu / m_s)( m_d / m_e) \approx 10.7^{+ 1.8}_{-0.8}$ \cite{Antusch:2013jca}, which is in tension with our assumption $(m_\mu / m_s)( m_d / m_e) = 9$ at the level of a few sigma.

Seesaw simplicity suggests that the two large angles in the PMNS lepton mixing matrix arise from a bi-large mixing matrix, e.g. TBM, assuming that the small reactor angle is entirely generated by Cabibbo haze from the charged leptons.

However, symmetric $Y^{(-\frac{1}{3})}$ textures in $SU(5)$ are incompatible with TBM mixing \cite{kile}, and ``seesaw simplicity'' requires us to search for the minimal asymmetry in $Y^{(-\frac{1}{3})}$ that yields the PMNS angles \cite{rrx2018asymmetric}. Under the assumption that all Yukawa couplings are real, and there is only one $\overline{\g{45}}$ coupling (inspired by minimality), a unique Georgi-Jarlskog-like \cite{gj} texture at the GUT scale emerges \cite{rrx2018asymmetric}:
\begin{align}
\begin{aligned}
Y^{(\frac{2}{3})} \sim~ &\mathrm{diag}\ (\lambda^8, \lambda^4, 1), \\
Y^{(-{1\over3})} \sim 
\begin{pmatrix}
 b d \lambda ^4 & a \lambda ^3 & b \lambda ^3 \cr
 a \lambda ^3 & c \lambda ^2 & g \lambda ^2 \cr
 d \lambda  & g \lambda ^2 & 1 \end{pmatrix}
~~&\mathrm{ and~~~}
Y^{(-1)} \sim 
\begin{pmatrix}
 b d \lambda ^4 & a \lambda ^3 & d \lambda \cr
 a \lambda ^3 & -3c \lambda ^2 & g \lambda ^2 \cr
 b \lambda^3  & g \lambda ^2 & 1 \end{pmatrix}, \label{texture}
\end{aligned}
\end{align}
where in terms of the Wolfenstein parameters $A$, $\rho$, and $\eta$, the $\mc O(1)$ prefactors \cite{rrx2018asymmetric} are \footnote{Curiously, there are numerical coincidences between prefactors: $\frac{2}{3A} \approx A = 0.81$ implying $g \approx d$, and $b = 0.31$ is close to $a = c = 0.33$.}
\begin{equation} \nonumber
a = c = \frac{1}{3}, ~~ g = A, ~~b = A \sqrt{\rho^2+\eta^2}, ~~d = \frac{2a}{g}=\frac{2}{3A}.
\end{equation}
\blu{The solo $\overline{\g{45}}$ coupling $c$ appears only in the $(22)$ position, and implies that the subdeterminant with respect to it must vanish to satisfy Eq.~\eqref{detcond}.} This texture successfully reproduces the Cabibbo-Kobayashi-Maskawa (CKM) mixing angles, the Gatto relation, and the GUT-scale mass relations (up to an overall constant) \cite{rrx2018asymmetric}:
\begin{align}
\begin{aligned}
&m_b = 1.019, \quad m_d = 0.994\ \lambda^4/3,\quad m_s = 0.951\ \lambda^2/3, \\
&m_\tau = 1.019, \quad m_e = 0.912\ \lambda^4/9,\quad m_\mu = 1.036\ \lambda^2. \label{mymass}
\end{aligned}
\end{align} 
These yield $(m_\mu / m_s)( m_d / m_e) \approx 10.7$, in agreement with Ref.~\cite{Antusch:2013jca} but in slight disagreement with Georgi-Jarlskog's relations in $SU(5)$, and do not change the structure of the texture.

With TBM seesaw mixing, the texture of Eq.~\eqref{texture} slightly overestimates the reactor and solar angles and underestimates the atmospheric angle. All angles are brought within $1\sigma$ of their PDG fit \cite{tanabashi2018review} by introducing a single phase $\delta = 78^\circ$ \cite{rrx2018asymmetric} in the TBM matrix: 
\begin{equation} \label{tbmdelta}
\mc U_{ TBM}(\delta) = \begin{pmatrix} 1 & 0 & 0 \\ 
0 &  1 &  0 \\  0 & 0 &  e^{i \delta} 
\end{pmatrix} \cdot
\begin{pmatrix} \sqrt{\frac{2}{3}} & \frac{1}{\sqrt{3}} & 0 \\ 
-\frac{1}{\sqrt{6}} &  \frac{1}{\sqrt{3}} &  \frac{1}{\sqrt{2}} \\  \frac{1}{\sqrt{6}} & - \frac{1}{\sqrt{3}} &  \frac{1}{\sqrt{2}} 
\end{pmatrix}.
\end{equation}
It generates a $\cancel{CP}$ Dirac phase $|\delta_{CP}| = 0.68\pi$, yielding the Jarlskog-Greenberg invariant \cite{jarlskog1, *greenberg1} $|\mathcal{J}| = 0.028$, consistent with the current PDG estimate \cite{tanabashi2018review}.
%It generates a $\cancel{CP}$ Jarkskog-Greenberg invariant \cite{jarlskog1, *greenberg1} $|\mc J| = 0.028$, consistent with the current PDG central value \cite{tanabashi2018review}.

Although phenomenologically successful, the asymmetric TBM texture was constructed without concern for fine-tuning. Its key features are
\begin{itemize}
    \item an asymmetric term in $Y^{(-\frac{1}{3})}$ and $Y^{(-1)}$,
    \item a vanishing subdeterminant with respect to the $(22)$ element of $Y^{(-\frac{1}{3})}$ and $Y^{(-1)}$,
    \item a diagonal hierarchical $Y^{(\frac{2}{3})}$,
    \item TBM seesaw mixing with a phase.
\end{itemize}
These fine-tuned features become natural when they originate from a discrete family symmetry.

In Ref.~\cite{Perez:2019aqq}, we identified $\mc T_{13} = \mc Z_{13} \rtimes \mc Z_3$ as the smallest non-Abelian discrete subgroup of $SU(3)$ capable of reproducing the first two features. Assuming the fermions $F$ and $T$ transform as $\mc T_{13}$ triplets but the Higgs $H$ as a family singlet, the $\Delta I_w = \frac{1}{2}$ effective operators are at least of dimension five: $FTH \varphi$, constructed with gauge-singlet family-triplet and -antitriplet familons $\varphi$. These interactions are mediated by heavy messengers with vectorlike mass. The vacuum alignment of the familons are ``crystallographic,'' pointing towards the sides or face diagonals of a three-dimensional cube.  

$\mc T_{13}$ contains two different triplet representations required by the asymmetry. Considering $F\equiv(F_1, F_2, F_3) \sim (\gb{5},\g{3}_1)$ and $T\equiv(T_1, T_3, T_2) \sim (\g{10},\g{3}_2)$ under $SU(5) \times \mc T_{13}$, it labels each matrix element $F_i T_j$ of the texture with a  unique $\mc Z_{13}$ charge and thus separates out the asymmetric term. \myred{The vanishing of the $(22)$-subdeterminant is achieved naturally by coupling the operator $FTH$ to three familons, orthogonal in the vacuum. The first two, $\varphi$ and $\varphi'$, couple at dimension five and generate the $F_1 T_3$ and $F_3 T_3$ elements, respectively. The third familon, $\varphi''$ then couples to \emph{both} dimension-five operators, generating the $F_1 T_1$ and $F_3 T_1$ elements at dimension six; their $\mc T_{13}$ coupling structure and vacuum alignments then implement the requisite relation between the matrix elements $Y^{(-\frac{1}{3})}_{11} Y^{(-\frac{1}{3})}_{33} = Y^{(-\frac{1}{3})}_{13} Y^{(-\frac{1}{3})}_{31}$, irrespective of the coupling constants. }

%In Ref.~\cite{Perez:2019aqq}, we explained the asymmetric texture from an underlying family symmetry. We showed that the natural separation of the asymmetric term leads us to the nonabelian group $\mc T_{13} = \mc Z_{13} \rtimes \mc Z_{3}$. We then constructed an explicit model that yields the asymmetric texture and shows how the equality of the determinants of $Y^{(-\frac{1}{3})}$ and $Y^{(-1)}$ can be explained. Considering $F\equiv(F_1, F_2, F_3) \sim (\gb{5},\g{3}_1)$ and $T\equiv(T_1, T_3, T_2) \sim (\g{10},\g{3}_2)$ under $SU(5) \times \mc T_{13}$, the texture is constructed from dimension-five and -six operators of the form $FTH \varphi$ and $FTH\varphi \varphi'$, with gauge-singlet, family-triplet familons $\varphi$, $\varphi'$ and family-singlet Higgs $H$, where $H$ transforms as either a $\gb{5}$ or $\overline{\g{45}}$ under $SU(5)$. These higher dimensional operators are constructed from tree-level vertices involving heavy messenger fields $\Delta$ and $\Sigma$. The model requires only `simple' familon vacuum alignments like $(1,0,0)$ or $(1,1,0)$ or their permutations. The symmetry is extended to include a $\mc Z_{5}$ factor to prevent unwanted vertices at the tree level. 

\vspace{0.5cm}
\blu{In this paper we complete the $SU(5) \times \mc T_{13}$ model by implementing the last two features -- diagonal $Y^{(\frac{2}{3})}$ and complex-TBM seesaw mixing -- of the asymmetric texture. In the next section, we show how the hierarchical structure of the up-type quark matrix appears naturally in the $\mc T_{13}$ model. }

\section{$Y^{(\frac{2}{3})}$ Texture}\label{sec:5}
Assuming a family-singlet Higgs $\Higgs$, the up-type quark Yukawa matrix $Y^{(\frac{2}{3})}$ is constructed from terms like $T T \Higgs \varphi$, where $\varphi$ is a gauge-singlet $\mc T_{13}$ triplet or antitriplet familon (or combination of such familons) and $\Higgs$ is the complex conjugate of the field $H_{\gb{5}}$ that couples to $Y^{(-\frac{1}{3})}$ and $Y^{(-1)}$. In terms of $\mc T_{13}$ Clebsch-Gordan coefficients, the product $T \otimes T$ yields
\begin{align} \label{eq:TT}
    \matc{
    T_1\\
    T_3\\
    T_2
    }_{\g{3}_2} \otimes
    \matc{
    T_1\\
    T_3\\
    T_2
    }_{\g{3}_2} &\rightarrow 
    \matc{
    T_3 T_3 \\
    T_2 T_2 \\
    T_1 T_1
    }_{\gb{3}_1} \oplus 
    \matc{
    T_3 T_2 \\
    T_2 T_1 \\
    T_1 T_3
    }_{\gb{3}_2} \oplus
    \matc{
    T_2 T_3 \\
    T_1 T_2 \\
    T_3 T_1
    }_{\gb{3}_2} . 
\end{align}

With simple familon vacuum alignments, the hierarchical structure of $Y^{(\frac{2}{3})}$ suggests the operators
\begin{align*}
     TT\Higgs\varphi^{(t)}_{\g{3}_1} \qquad &\textrm{for the top-quark mass,}\\
     TT\Higgs\varphi^{(t)}_{\g{3}_1} \varphi_{\g{3}_i} \qquad &\textrm{for the charm-quark mass, and}\\
     TT\Higgs\varphi^{(t)}_{\g{3}_1} \varphi_{\g{3}_i} \varphi_{\g{3}_i} \qquad &\textrm{for the up-type quark mass}
\end{align*}
in vacuum, with the hierarchical factor of $\lambda^4$ supplied by $\vev{\varphi_{\g{3}_i}}$. $\varphi^{(t)}_{\g{3}_1}$ transforms as a $\g{3}_1$, while $\varphi_{\g{3}_i}$ is a triplet or antitriplet whose exact representation is unresolved at this stage. 

\subsection{Top-quark mass}
The dimension-five operator $TT\Higgs\varphi^{(t)}_{\g{3}_1}$ yields the top-quark mass  when $\vev{\varphi^{(t)}_{\g{3}_1}}  \sim m_t (1,0,0).$ It arises from  tree-level vertices $T\Gamma \vphi$ and $T\overline{\Gamma}\Higgs$, where $\Gamma \sim (\overline{\g{10}}, \g{3}_2)$ under $SU(5) \times \mc T_{13}$ is a heavy messenger field with   vectorlike mass:
\begin{equation}\label{dia:charm}
	\centering
	\begin{tikzpicture}[baseline=(a.base)]
	\begin{feynman}[small]
	\vertex (a);
	\vertex [right=of a] (c);
	\vertex [above left=of a] (i1) {\(T\)};
	\vertex [below left=of a] (i2) {\(\varphi^{(t)}_{\g{3}_1}\)};
	\vertex [above right=of c] (f1) {\(T\)};
	\vertex [below right=of c] (f2) {\(\bar{H}_{\g{5}}\)};
	\diagram* {
		(i1) -- (a) --[insertion=0.5, edge label=\(\Gamma\quad \overline{\Gamma}\)] (c) -- (f1),
		(a) -- [scalar] (i2),
		(c) -- [scalar] (f2),
	};
	\end{feynman}
	\end{tikzpicture}
	%\caption{Effective operator}	
\end{equation}
This diagram implements the contractions $(T \vphi)_{\gb{3}_2} \cdot (T H_\g{5})_{\g{3}_2}.$ The first contraction yields $m_t(0, T_3, 0)_{\gb{3}_2}$ and the second $(T_1, T_3, T_2)_{\g{3}_2}$, resulting in the top-quark mass term $  m_t T_3 T_3$ \footnote{The dimension-five operator generating the top-quark mass is the leading order (LO) operator. As will be discussed in Section \ref{sec:6} and Appendix \ref{app:shape}, this is guaranteed by introducing a $\mathcal{Z}_n$ symmetry to restrict `dangerous' operators from coupling to unwanted matrix elements. Even without the $\mathcal{Z}_n$ symmetry, the next-to-leading-order (NLO) operator contributing to the top-quark mass is of dimension six: $TT\bar{H}_{\textbf{5}}\varphi_{3_1}^{(t)*} \varphi_{3_2}$, which is suppressed  by a factor of $\lambda^4$.}. 

\subsection{Charm-quark mass}
From Eq. \eqref{eq:TT}, we want the familon combination $\vphi \varphi_{\g{3}_i}$ to transform as a $\g{3}_1$, with a vacuum alignment along $(0,1,0)$. The $\mc T_{13}$ Kronecker products then uniquely determine $\varphi_{\g{3}_i} \equiv \varphi_{\g{3}_2}$. 

The dimension-six operator $TT\Higgs \vphi \varphi_{\g{3}_2}$ can be constructed by adding two new tree-level vertices $T\Omega\vphi$ and $\overline{\Omega}\Gamma \varphi_{\g{3}_2}$ to $T\overline{\Gamma}\Higgs$: 
\begin{equation}
\begin{tikzpicture}[baseline=(a.base)]
\begin{feynman}[small]
\vertex (a);
\vertex [right=of a] (b);
\vertex [right=of b] (c);
\vertex [below=of b] (d) {$\varphi_{\g{3}_2}$};
\vertex [above left=of a] (i1) {\(T\)};
\vertex [below left=of a] (i2) {\(\vphi\)};
\vertex [above right=of c] (f1)	 {\(T\)};
\vertex [below right=of c] (f2) {\(\bar{H}_{\g{5}}\)};
\diagram* {
	(i1) -- (a) --[insertion=0.5, edge label=\(\Omega\ \overline{\Omega}\)] (b) --[insertion=0.5, edge label=\(\Gamma\ \overline{\Gamma}\)] (c) -- (f1),
	(a) -- [scalar] (i2),
	(b) -- [scalar] (d),
	(c) -- [scalar] (f2),
};
\end{feynman}
\end{tikzpicture}
\end{equation}
giving the contraction
$\left(T \vphi\right)_{\g{3}_2} \cdot \left(TH\varphi_{\g{3}_2}\right)_{\gb{3}_2}$,
where a new vectorlike messenger $\Omega \sim (\overline{\g{10}}, \gb{3}_2)$ is required to pick out $m_t T_2$ from the first contraction. With $\vev{\varphi_{\g{3}_2}} \sim \lambda^4 (1, \alpha, 0)$, where $\alpha$ is still unresolved, the second contraction contributes $\lambda^4 T_2$, thus resulting in the charm-quark mass term $m_t \lambda^4 T_2 T_2$ in vacuum.

\subsection{Up-quark mass}
Again consulting Eq. \eqref{eq:TT}, the familon combination $\vphi \varphi_{\g{3}_2}\varphi_{\g{3}_2}$ must transform as a $\g{3}_1$ and be aligned along $(0,0,1)$ in vacuum. This fixes $\alpha$ to be $1$.

The dimension-seven operator $TT\Higgs \vphi \varphi_{\g{3}_2}\varphi_{\g{3}_2}$ can be constructed by adding three new tree-level vertices $T\Theta\vphi$, $\overline{\Theta}\Theta\varphi_{\g{3}_2}$ and $\overline{\Theta}\Gamma \varphi_{\g{3}_2}$ to $T\overline{\Gamma}\Higgs$, where $\Theta \sim (\overline{\g{10}}, \gb{3}_1)$ is a new messenger:
\begin{equation}\label{dia:up}
\begin{tikzpicture}[baseline=(a.base)]
\begin{feynman}[small]
\vertex (a);
\vertex [right=of a] (b);
\vertex [right=of b] (c);
\vertex [right=of c] (d);
\vertex [below=of b] (e){$\varphi_{\g{3}_2}$};
\vertex [below=of c] (f){$\varphi_{\g{3}_2}$};
\vertex [above left=of a] (i1) {\(T\)};
\vertex [below left=of a] (i2) {\(\vphi\)};
\vertex [above right=of d] (f1)	 {\(T\)};
\vertex [below right=of d] (f2) {\(\bar{H}_{\g{5}}\)};
\diagram* {
	(i1) -- (a) --[insertion=0.5, edge label=\(\Theta\ \overline{\Theta}\)] (b) --[insertion=0.5, edge label=\(\Theta\ \overline{\Theta}\)] (c) --[insertion=0.5, edge label=\(\Gamma\ \overline{\Gamma}\)] (d) -- (f1),
	(a) -- [scalar] (i2),
	(b) -- [scalar] (e),
	(c) -- [scalar] (f),
	(d) -- [scalar] (f2),
};
\end{feynman}
\end{tikzpicture}
\end{equation}
implementing the contractions
$
    \left( (T\vphi)_{\g{3}_1} \cdot \varphi_{\g{3}_2}\right)_{\g{3}_1} \cdot \left( TH_\g{5} \varphi_{\g{3}_2} \right)_{\gb{3}_1}.
$
The first contraction extracts $m_t T_1$, the second $\lambda^4$, while the third gives $\lambda^4 T_1$, thus yielding the up- quark mass term $m_t \lambda^8 T_1 T_1$ in vacuum. 

\myred{In summary, the above diagrams yield the desired hierarchical up-type quark masses:}
\begin{align*}
    m_u : m_c : m_t = \lambda^8 : \lambda^4 : 1. 
\end{align*}

%This completes our discussion of the $Y^{(\frac{2}{3})}$ texture. In the next section, we focus on the seesaw sector and show how TBM-diagonalization of the seesaw matrix and prediction for light-neutrino masses are obtained.

\section{The seesaw Sector}\label{sec:3}
\blu{In this section we show how TBM seesaw mixing is realized in the $SU(5) \times \mc T_{13}$ model. It requires four right-handed neutrinos and three familons, whose  vacuum expectation values need not be fine-tuned to yield TBM-diagonalization.}

 The necessity of the fourth right-handed neutrino becomes apparent by first considering the simpler three-neutrino case.

\subsection{Three Right-Handed Neutrinos}
We introduce three right-handed neutrinos  $\bar{N} \equiv (\bar{N}_1, \bar{N}_3, \bar{N}_2)$, \myred{their order mimicking $T \equiv (T_1, T_3, T_2)$ inspired by an $SO(10)$ extension of the gauge group, and} transforming as $(\g{1}, \g{3}_2)$ under $SU(5) \times \mc T_{13}$.
%\footnote{\blu{This is inspired by contemplating an $SO(10)$ extension of the gauge group. In an $SO(10) \times \mc T_{13}$ model, the $(\gb{5},\g{3}_1)$ and $(\g{10},\g{3}_2)$ of $SU(5)\times \mc T_{13}$ are embedded in $(\g{10},\g{3}_1)$ and $(\g{16},\g{3}_2)$:
%\begin{align*}
%    (\g{10},\g{3}_1) \oplus (\g{16},\g{3}_2) \rightarrow  \left((\g{5},\g{3}_1) \oplus (\gb{5},\g{3}_1)\right) \oplus \left( (\gb{5},\g{3}_2) \oplus (\g{10},\g{3}_2) \oplus (\g{1},\g{3}_2)\right)  
%\end{align*}
%We identify the $(\g{1},\g{3}_2)$ as the field $\bar{N}$ and label it similar to the field $T \equiv (T_1, T_3, T_2) \sim (\g{10},\g{3}_2)$.}}
Their $\Delta I_w = \frac{1}{2}$ coupling is given by the dimension-five operator 
$
F \bar{N} \bar{H}_{\g{5}} \varphi_{\mc A},
$
where $\varphi_{\mc A}$ is a familon transforming as $(\g{1}, \gb{3}_1 \times \gb{3}_2) = (\g{1}, \gb{3}_1) \oplus (\g{1}, \gb{3}_2) \oplus (\g{1}, \g{3}_2)$. This operator can be constructed from tree-level vertices $y_{\mc A} F \Lambda \bar{H}_{\g{5}}$ and $y_{\mc A}' \bar{N} \overline{\Lambda}\varphi_{\mc A}$:

\begin{equation}
\begin{comment}
\begin{tikzpicture}[baseline=(a.base)]
    \begin{feynman}[small]
        \vertex (a);
        \vertex [right=of a] (c){\({\Lambda}\)};
        \vertex [above left=of a] (i1) {\(F\)};
        \vertex [below left=of a] (i2) {\(\bar{H}_{\g{5}}\)};
        \diagram* {
        	(i1) -- (a) -- (c),
        	(a) -- [scalar] (i2),
        };
    \end{feynman}
\end{tikzpicture} + 
\begin{tikzpicture}[baseline=(a.base)]
    \begin{feynman}[small]
        \vertex (a);
        \vertex [left=of a] (c){\({\overline{\Lambda}}\)};
        \vertex [above right=of a] (i1) {\(\bar{N}\)};
        \vertex [below right=of a] (i2) {\(\varphi_{\mc A}\)};
        \diagram* {
        	(i1) -- (a) -- (c),
        	(a) -- [scalar] (i2),
        };
    \end{feynman}
\end{tikzpicture} \rightarrow
\end{comment}
\begin{tikzpicture}[baseline=(a.base)]
	\begin{feynman}[small]
	\vertex (a);
	\vertex [right=of a] (c);
	\vertex [above left=of a] (i1) {\(F\)};
	\vertex [below left=of a] (i2) {\(\bar{H}_{\g{5}}\)};
	\vertex [above right=of c] (f1) {\(\bar{N}\)};
	\vertex [below right=of c] (f2) {\(\varphi_{\mc A}\)};
	\diagram* {
		(i1) -- (a) --[insertion=0.5, edge label=\(\Lambda\quad \overline{\Lambda}\)] (c) -- (f1),
		(a) -- [scalar] (i2),
		(c) -- [scalar] (f2),
	};
	\end{feynman}
\end{tikzpicture}
	\rightarrow
	\frac{1}{M}y_{\mc A} y_{\mc A}' \vev{\bar{H}_{\g{5}}} \vev{\varphi_{\mc A}} F\bar{N}.	
	\label{feynman}
\end{equation}
Here, $y_{\mc A}$ and $y_{\mc A}'$ are dimensionless Yukawa couplings and $\Lambda$ is a complex messenger with heavy vectorlike mass $M$. Denoting the combination of vacuum expectation values of the familon and Higgs as
$\frac{1}{M}y_{\mc A} y_{\mc A}' \vev{\bar{H}_{\g{5}}} \vev{\varphi_{\mc A}} \equiv (a_1, a_2, a_3)^t,$
$\mc T_{13}$ yields three possibilities for the coupling matrix $\mc A$:
\begin{align}
\varphi_{\mc A} \sim \gb{3}_1: \left(
			\begin{array}{ccc}
 			a_2 & 0 & 0 \\
 			0 & 0 & a_1 \\
 			0 & a_3 & 0 \\
			\end{array}
		\right),\ 
\varphi_{\mc A} \sim \gb{3}_2: \left(
			\begin{array}{ccc}
 			0 & a_3 & 0 \\
 			a_2 & 0 & 0 \\
 			0 & 0 & a_1 \\
			\end{array}
		\right),
\varphi_{\mc A} \sim \g{3}_2: \left(
			\begin{array}{ccc}
 			0 & 0 & a_1 \\
 			0 & a_3 & 0 \\
 			a_2 & 0 & 0 \\
			\end{array}
		\right), \label{Aforms2}
\end{align}
where the $a_i$ have dimension of mass.

The $\Delta I_w = 0$ coupling of the right-handed neutrinos is given minimally by the dimension-four operator
$
y_{\mc B}\bar{N} \bar{N} \varphi_{\mc B}
$
for some dimensionless coupling constant $y_{\mc B}$, where $\varphi_{\mc B}$ transforms as $(\g{1},\gb{3}_2 \times \gb{3}_2) = (\g{1},\g{3}_1) \oplus (\g{1},\g{3}_2) \oplus (\g{1},\g{3}_2)$. Denoting its vacuum expectation value by $y_{\mc B} \vev{\varphi_{\mc B}} \equiv (b_1, b_2, b_3)^t$, $\mc T_{13}$ offers two possibilities for the symmetric Majorana matrix $\mc B$:
\begin{align}
\varphi_{\mc B} \sim \g{3}_2: \left(
			\begin{array}{ccc}
 			0 & b_2 & b_3 \\
 			b_2 & 0 & b_1 \\
 			b_3 & b_1 & 0 \\
			\end{array}
		\right), \quad
\varphi_{\mc B} \sim \g{3}_1: \left(
			\begin{array}{ccc}
 			b_3 & 0 & 0 \\
 			0 & b_2 & 0 \\
 			0 & 0 & b_1 \\
			\end{array}
		\right), \label{Bforms2}
\end{align} 
where again the $b_i$ have dimension of mass. 

%In general, there could be a linear combination of familons contributing to $\mc A$ and $\mc B$, but we consider only one familon.
\myred{Minimality dictates we introduce the least number of right-handed neutrinos and familons in the seesaw sector. In this spirit, we adopt $\varphi_{\mc A} \sim \gb{3}_2$ in Eq.~\eqref{Aforms2} and $\varphi_{\mc B} \sim \g{3}_2$ in Eq.~\eqref{Bforms2}. The implications of the alternative choices are discussed in Appendix~\ref{App:B}.}

\vspace{0.5cm}
The seesaw matrix $\mc S$ is related to the $\Delta I_w = \frac{1}{2}$ and $\Delta I_w = 0$ matrices by
\begin{align}
\mc S = \mc A \mc B^{-1} \mc A^t,
\end{align}
for $\det \mc B \neq  0$. We choose a particular decomposition of $\mc B$:
\begin{align}
\mc B = \mc C\ \mc G\ \mc C^t, \label{mydecom}
\end{align}
where $\mc C$ depends on $\vev{\varphi_{\mc B}}$: 
%%%%%%%%%%%%%%%%%%%
\begin{comment}
Compare this to the case of Golden Ratio mixing. By choosing $\mc C = \sqrt{b_1 b_2 b_3}\ \text{diag}(b_1^{-1}, \sqrt{2} b_3^{-1}, -b_2^{-1})$, the decomposition of Eq.~\eqref{mydecom} yields 
\begin{align*}
    \mc G = \left(
			\begin{array}{ccc}
 			0 & \frac{1}{\sqrt{2}} & -\frac{1}{\sqrt{2}} \\
 			\frac{1}{\sqrt{2}} & 0 & -1 \\
 			-\frac{1}{\sqrt{2}} & -1 & 0 \\
			\end{array}
		\right) \label{Gform}
\end{align*}
which is diagonalized by the Golden Ratio matrix. In this case, however, $\mc C$ is not proportional to a unitary matrix without fine-tuning the vacuum expectation value of $\varphi_{\mc B}$ and it cannot be absorbed into the redefinition of $\bar{N}$. Diagonalizing the seesaw matrix matrix by the Golden Ratio mixing does not follow naturally, it requires $vev{\varphi_{\mc A}}$ and  $vev{\varphi_{\mc B}}$ to be related in a particular way.
\end{comment}
%%%%%%%%%%%%%%%%%%%%%%%%%%
\begin{align}
\mc C = \sqrt{b_1 b_2 b_3}\left(
			\begin{array}{ccc}
 			b_1^{-1} & 0 & 0 \\
 			0 & b_3^{-1} & 0 \\
 			0 & 0 & -b_2^{-1} \\
			\end{array}
		\right), 
\end{align}
and $\mc G$ is a purely numerical matrix:
\begin{align}
\mc G = \left(
			\begin{array}{ccc}
 			0 & 1 & -1 \\
 			1 & 0 & -1 \\
 			-1 & -1 & 0 \\
			\end{array}
		\right). \label{Gform}
\end{align}
%\footnote{\blu{This identification of $\mc C$ and $\mc G$ is not unique; one can write $\mc B = \mc C'\ \mc G'\ \mc C'^t$, where $\mc C' = \mc C\ \mc V$ and $\mc G' = \mc V^{-1}\ \mc G\ (\mc V^t)^{-1}$ for some arbitrary invertible matrix $\mc V$. We proceed with the identification in Eq.~\eqref{CGform} because it leads to TBM-seesaw mixing, as we will show.}}
Surprisingly, $\mc G$ is diagonalized by the TBM matrix
\begin{align}
\mc G = \mc U_{TBM}\ \mc D_b\ \mc U_{TBM}^t, \label{Gdecomp}
\end{align}
where
$
\mc D_b = \text{diag} (-1, 2, -1).
$
$\mc G$ is invariant under the transformation $\mc P'$ 
\begin{align}
\mc P'\ \mc G\ \mc P'^t = \mc G, \label{Ptrans}
\end{align}
so that $\mc C$ can be redefined as 
\begin{align}
\mc C \rightarrow \mc C\ \mc P' \label{ctrans}
\end{align}
in Eq.~\eqref{mydecom}, where $\mc P'$ is the identity matrix or any of the following permutation  matrices (up to a sign)
\begin{align}
\begin{gathered}
(1\ 2):\left(
			\begin{array}{ccc}
 			0 & 1 & 0 \\
 			1 & 0 & 0 \\
 			0 & 0 & 1 \\
			\end{array}
		\right),\ 
(2\ 3):\left(
			\begin{array}{ccc}
 			-1 & 0 & 0 \\
 			0 & 0 & 1 \\
 			0 & 1 & 0 \\
			\end{array}
		\right),\ 
(3\ 1):\left(
			\begin{array}{ccc}
 			 0 & 0 & 1 \\
 			0 & -1 & 0 \\
 			1 & 0 & 0 \\
			\end{array}
		\right), \\
(1\ 2\ 3):\left(
			\begin{array}{ccc}
 			0 & -1 & 0 \\
 			0 & 0 & 1 \\
 			1 & 0 & 0 \\
			\end{array}
		\right),\
(3\ 2\ 1):\left(
			\begin{array}{ccc}
 			0 & 0 & 1 \\
 			-1 & 0 & 0 \\
 			0 & 1 & 0 \\
			\end{array}
		\right).  \label{Pforms2}
\end{gathered}
\end{align}

Using the decomposition of Eq.~\eqref{mydecom}, the seesaw matrix is given by 
\begin{align}
\mc S = \mc A (\mc C^{-1})^t\ \mc U_{TBM}\ \mc D_b^{-1}\ \mc U_{TBM}^t\ \mc C^{-1} \mc A^t. \label{seesaw}
\end{align}
$\mc S$ is itself diagonalized by $\mc U_{TBM}(\delta)$ only if 
\begin{align}
\mc A (\mc C^t)^{-1} &= \sqrt{m_\nu}\ \text{diag}\ (1,1,e^{i\delta})\ \mc P'^t \nonumber \\
 \implies \mc A &= \sqrt{m_\nu}\ \text{diag}\ (1,1,e^{i\delta})\ (\mc C \mc P')^t           \label{cond1}
\end{align}
for some mass parameter $m_\nu$.

Eq.~\eqref{cond1} embodies two requirements: (i) $\mc A$ must have the same form as $(\mc C \mc P')^t$, and (ii) the vacuum alignment of $\varphi_{\mc A}$, given by $a_i$, is determined by that of $\varphi_{\mc B}$, given by $b_i$. 

\noindent Requirement (i) can always be satisfied; for any $\mc A$ in \eqref{Aforms2}, there exists a $\mc P'$ in \eqref{Pforms2} that satisfies Eq.~\eqref{cond1}. With $\varphi_{\mc A} \sim \gb{3}_2$ and $\mc P' \equiv (1\ 2)$, we have 
\begin{align}
\mc C \rightarrow \mc C \mc P' = \sqrt{b_1 b_2 b_3} \left(
			\begin{array}{ccc}
 			0 & b_1^{-1} & 0 \\
 			b_3^{-1} & 0 & 0 \\
 			0 & 0 & -b_2^{-1} \\
			\end{array}
		\right), \label{Cform}
\end{align}
yielding the same $\mc G$ as in Eq.~\eqref{Gform}. With this form of $\mc C$, requirement (ii) is fulfilled by the alignment
\begin{align}
\matc{a_1\\ a_2\\ a_3} =  \sqrt{m_\nu b_1 b_2 b_3} \matc{-b_2^{-1} e^{i\delta}\\ b_1^{-1}\\ b_3^{-1}}. \label{avev}
\end{align}

\begin{comment}
This suggests that in vacuum, $\varphi_{\mc A}$ can be expressed as a composite familon made up of two $\varphi_{\mc B}$s:
\begin{align}
    \vev{\varphi_{\mc A}} &= \sqrt{\frac{m_\nu}{b_1 b_2 b_3}} \left(
			\begin{array}{ccc}
 			0 & -e^{i\delta} & 0 \\
 			1 & 0 & 0 \\
 			0 & 0 & 1 \\
			\end{array}
		\right) \vev{\left(\varphi_{\mc B} \cdot \varphi_{\mc B}\right)_{\gb{3}_2}} .
\end{align}
\end{comment}

Applying Eq.~\eqref{cond1}, the seesaw matrix becomes
\begin{align}
\mc S = m_\nu\ \mc U_{TBM}(\delta)\ \text{diag} \left(-1,\frac{1}{2},-1 \right)\ \mc U_{TBM}^t(\delta)
\end{align} 
\myred{and yields three relations among the light neutrino masses: $$m_{\nu_2} = \frac{1}{2}m_{\nu_3}, \quad  m_{\nu_1} = 2 m_{\nu_2},\quad \text{and\quad } m_{\nu_1} = m_{\nu_3}.$$ The first relation is consistent with normal ordering, but the other two, involving $m_{\nu_1}$, contradict oscillation data \cite{tanabashi2018review}. If $m_{\nu_1}$ can be corrected to a smaller value, the first relation can be used along with oscillation data to calculate the light neutrino masses in normal ordering. We are then compelled to enlarge the neutrino sector.} 

%\blu{The analysis of the three right-handed neutrino case shows that TBM-seesaw mixing arises from the off-diagonal symmetric structure of $\mc B$, dictated by $\mc T_{13}$ Clebsch-Gordan coefficients, and its decomposition in Eqs.~\eqref{mydecom}, \eqref{CGform} and \eqref{Gdecomp}. This construction rules out inverted ordering, but would predict normal ordering if $m_{\nu_1}$ were smaller.  The minimal extension is to introduce a fourth right-handed neutrino, which is expected to correct $m_{\nu_1}$.}

%\blue{We are then compelled to extend the neutrino sector by introducing a fourth right-handed neutrino.}

\subsection{Four Right-Handed Neutrinos}
Following our minimalist approach, we choose a gauge- and family-singlet fourth right-handed neutrino $\bar{N}_4$. It introduces the extra operators
\begin{align*}
\Delta I_w = \frac{1}{2}:\quad &   F\bar{N}_4 \bar{H}_{\g{5}} \varphi_v, \quad \text{where } \varphi_{v} \sim (\g{1},\gb{3}_1), \\
\Delta I_w = 0:\quad &y_z \bar{N} \bar{N_4} \varphi_z,\ \text{and } m\bar{N}_4 \bar{N_4}, \quad \text{where } \varphi_z \sim (\g{1},\gb{3}_2).
\end{align*}
The dimension-five operator $F\bar{N}_4 \bar{H}_{\g{5}} \varphi_v$ can be constructed from tree-level vertices $y_{\mc A} F \Lambda \Higgs$ and $y_v' \bar{N} \overline{\Lambda} \varphi_v$ in a similar way as in Eq.~\eqref{feynman}, using the same messenger field $\Lambda$. For $\frac{1}{M}y_v y_v' \vev{\bar{H}_{\g{5}}} \vev{\varphi_v} \equiv v \equiv (v_1, v_2, v_3)^t$, the numerator of the seesaw formula is a $(3\times 4)$ $\Delta I_w = \frac{1}{2}$ matrix $$\matc{\mc A\quad v}$$.

The $(4\times 4)$ $\Delta I_w = 0$  Majorana matrix in vacuum is given by
\begin{align}
\mc M = \left(
\begin{array}{cc}
 \mc B & z \\
 z^t & m \\
\end{array}
\right),
\end{align}
where $y_z \vev{\varphi_z} \equiv z \equiv (z_1, z_2, z_3)^t$. For $\det \mc B \neq 0$, 
\begin{align}
\mc M^{-1} &= \left(
				\begin{array}{cc}
 					\mc B & z \\
 					z^t & m \\
				\end{array}
			  \right)^{-1} = \frac{1}{\mu}
			  \left(
				\begin{array}{cc}
 					\mu \mc B^{-1} + \mc B^{-1} z z^t \mc B^{-1}  & -\mc B^{-1} z \\
 					-z^t \mc B^{-1} & 1 \\
				\end{array}
			  \right),
\end{align}
with 
\begin{align}
    \mu = m - z^t \mc B^{-1} z. \label{mudef}
\end{align}

The seesaw matrix now has two terms:
\begin{align}
\mc S &\equiv \mc S_1 + \mc S_2 \nonumber \\
&= \mc A \mc B^{-1} \mc A^t + \frac{1}{\mu} \mc W\ \mc W^t , \label{mastereq}
\end{align}
where $$\mc W=\mc A \mc B^{-1} z-v.$$ The first term is the same as in the three right-handed neutrinos case:
\begin{align}
\mc S_1 &= m_\nu\ \mc U_{TBM} (\delta)\ \text{diag}\left(-1,\frac{1}{2},-1\right)\ \mc U_{TBM}^t (\delta).
\end{align} 

\noindent The second term  $\mc S_2= \frac{1}{\mu} \mc W\ \mc W^t$ has two zero eigenvalues. If it is to be diagonalized by $\mc U_{TBM} (\delta)$, the column vector $\mc U_{TBM}^\dagger (\delta) \mc W$ must be one  of the following forms:
$$
(0, 1, 0)^t,\quad   (0,0,1)^t,\quad  (1,0,0)^t.
$$

\noindent The first two are incompatible with data. A nonzero entry in the second element implies that $\mc S_2$ corrects only $m_{\nu_2}$,  leaving $m_{\nu_1}$ and $m_{\nu_3}$ degenerate. The third nonzero element is also unphysical because it leads to $m_{\nu_1} > m_{\nu_2}$.

Phenomenology requires us to choose the third possibility, in which case $\mc W$ is of the form 
\begin{equation}\label{w100}
    \mc W \propto \mc U_{TBM} (\delta)\matc{1\\ 0\\ 0} \propto \matc{2\\ -1\\ e^{i\delta}},
\end{equation}
\myred{which further aligns $\varphi_z$, $\varphi_v$ and $\varphi_{\mc B}$ in vacuum and corrects $m_{\nu_1}$. Thus $\mc S_2$ negates the two unwanted mass relations in $\mc S_1$, but the relation $m_{\nu_2} = \frac{1}{2} m_{\nu_3}$ singling out normal ordering remains unaltered. Together with oscillation data, it can determine all three light neutrino masses.}  

\blu{We present two minimal scenarios with either $\varphi_v$ or $\varphi_z$ absent in the seesaw formula. Both scenarios yield the same light neutrino mass spectrum.}

\subsection*{\textbf{\large Scenario 1: $\varphi_{\mc B} \sim \g{3}_2, \varphi_z \sim \gb{3}_2, \varphi_{\mc A} \sim \gb{3}_2$}}
In this case $\varphi_v$ is absent, and $\mc W=\mc A \mc B^{-1} z$. Applying Eqs.~\eqref{cond1} and \eqref{Gdecomp}, we obtain
\begin{align}
\mc W &=\sqrt{m_\nu}\ \mc U_{TBM}(\delta)\ \mc D_b^{-1}\ \mc U_{TBM}^t\ \mc C^{-1} z.
\end{align}

\noindent For $\mc C$ given by Eq.~\eqref{Cform}, it becomes
\begin{align}
 \mc W &= \frac{\sqrt{m_\nu}}{\sqrt{6b_1 b_2 b_3}}\ \mc U_{TBM}(\delta)\ \matc{(b_1 z_1 - 2b_3 z_2 + b_2 z_3)\\ \frac{1}{\sqrt{2}}(b_1 z_1 + b_2z_3 + b_3z_2)\\ -\sqrt{3}(b_1z_1 - b_2 z_3)}.
\end{align}
Comparing this to  Eq.~\eqref{w100}, we require
\begin{align}
b_1 z_1 + b_2z_3 + b_3z_2 = 0,\quad
b_1z_1 - b_2 z_3 = 0.
\end{align}
These constraints yield a vacuum alignment condition between $\varphi_{\mc B}$ and $\varphi_{z}$:
\begin{align}
b_1 z_1 = b_2 z_3 = -\frac{1}{2}b_3 z_2 \equiv m_{bz}^2  \label{bzsol}
\end{align}
where the parameter $m_{bz}$ has dimension of mass. Then $\vev{\varphi_z}$ becomes
\begin{align}
\matc{z_1\\ z_2\\ z_3} &= m_{bz}^2 \matc{b_1^{-1}\\ -2b_3^{-1}\\ b_2^{-1}}. \label{zvev}
\end{align}

\begin{comment}
which implies that $\varphi_z$, in vacuum, can be expressed as a composite familon made up of two $\varphi_{\mc B}$s:
\begin{align}
    \vev{\varphi_z} &= \frac{m_{bz}^2}{b_1 b_2 b_3} \left(
			\begin{array}{ccc}
 			1 & 0 & 0 \\
 			0 & 0 & -2 \\
 			0 & 1 & 0 \\
			\end{array}
		\right) \vev{\left(\varphi_{\mc B} \cdot \varphi_{\mc B}\right)_{\gb{3}_2}}.
\end{align}
\end{comment}
\noindent From Eq.~\eqref{mudef}, $\mu$ is evaluated as
\begin{align}
\mu &= \frac{6 m_{bz}^4+mb_1b_2b_3}{b_1b_2b_3}, \label{muval}
\end{align}
and $\mc S_2$ becomes
\begin{align}
\mc S_2= \frac{6 m_{\nu} m_{bz}^4}{6m_{bz}^4 + mb_1b_2b_3}\ \mc U_{TBM}(\delta)\ \text{diag}(1,0,0)\ \mc U_{TBM}^t(\delta).
\end{align}

%\vspace{0.3cm}
\noindent Combining $\mc S_1$ and $\mc S_2$ yields the light neutrino masses in normal ordering:
\begin{align}
m_{\nu_1} = -\frac{m}{\mu}\ m_{\nu}, \quad
m_{\nu_2} = \frac{1}{2}\ m_{\nu},\quad
m_{\nu_3} = - m_{\nu}, \label{numass3}
\end{align}
in terms of three undetermined parameters $m$, $\mu$ and $m_{\nu}$. As we will show below, $\frac{m}{\mu}$ and $m_{\nu}$ can be extracted from oscillation data, albeit with a sign ambiguity.  

\subsection*{\textbf{\large A circle parametrization for neutrino oscillations}}
\begin{figure}[!h]
\centering
\begin{tikzpicture}
\draw (0,0) circle [radius=2.3];
\coordinate (A) at (-2.3,0);
\coordinate (C) at (2.3,0);
\coordinate (B) at (50:2.3);
\coordinate (D) at (-60:2.3);
\draw (A)node[left]{$A$} --node[above left]{$\Delta_{31}$} (B)node[above right]{$B$} --node[below left]{$m_{\nu_1}$} (C) node[right]{$C$}--node[above left]{$m_{\nu_2}$} (D)node[below right]{$D$} --node[below left]{$\Delta_{32}$} (A) --node[above]{$m_{\nu_3}$} (C);
\end{tikzpicture}
\caption{Circle parametrization of neutrino masses and oscillation parameters.} \label{fig:numass}
\end{figure}
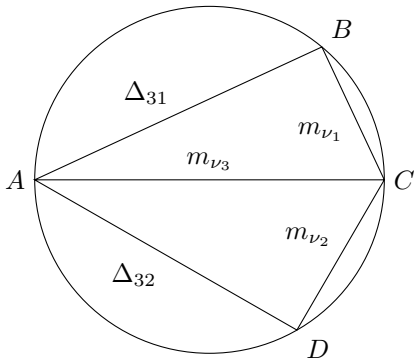

We introduce a convenient geometrical representation of oscillation parameters and neutrino masses. The neutrino oscillation parameters for normal ordering $\Delta_{31} \equiv \sqrt{\Delta m_{31}^2}$ and $\Delta_{32} \equiv \sqrt{\Delta m_{32}^2}$ and the light neutrino masses $m_{\nu_1}$, $m_{\nu_2}$ and $m_{\nu_3}$ are represented as the sides and diagonals of the inscribed quadrilateral $ABCD$ in Fig.~\ref{fig:numass}. The largest mass, $m_{\nu_3}$, is chosen to be the diameter of the circle. 

%\vspace{0.5cm}

%\vspace{0.3cm}
The relation $m_{\nu_2} = \frac{1}{2} m_{\nu_3}$ implies $\angle CAD = 30^\circ$. Using PDG values \cite{tanabashi2018review} of the oscillation data (see \cite{deSalas:2018bym, *Esteban:2018azc} for other recent global fits) for normal ordering,
%\begin{align}
%\Delta m_{21}^2 = 7.53 \times 10^{-5}\ eV^2,\quad \Delta m_{32}^2 = 2.51 \times 10^{-3}\ eV^2, \label{osciparam}
%\end{align}
we find
\begin{align*}
\Delta_{31} &= \sqrt{\Delta m_{32}^2 + \Delta m_{21}^2} = 50.8\ \text{meV}, \\
\Delta_{32} &= \sqrt{\Delta m_{32}^2} = 50.1\ \text{meV}. 
\end{align*}

\noindent Our prediction for the light neutrino masses follow:
\begin{align}
\boxed{
m_{\nu_1} = 27.6\ \text{meV},\quad m_{\nu_2} = 28.9\ \text{meV}, \quad m_{\nu_3} = 57.8\ \text{meV}. \label{numass2}
}
\end{align} 
\blu{Their sum is $114.3$ $\text{meV}$, very close to Planck's cosmological upper bound \cite{aghanim2018planck}}
\begin{align*}
    \sum_i |m_{\nu_i}| \leq 120\ \mathrm{meV}.
\end{align*}
%\myred{Our model can therefore be tested against data with a little improvement of this bound.}

\vspace{0.3cm}
\noindent Comparing Eqs.~\eqref{numass3} and \eqref{numass2}, the parameters $m_{\nu}$, $m$ and $\mu$ are given by
\begin{align}
|m_\nu| = 57.8\ \mathrm{meV}, \quad \left|\frac{m}{\mu}\right| = 0.48. \label{param1}
\end{align}
The sign ambiguity appears because these are determined from mass-squared relations in the oscillation data.
%\blue{So far we have introduced some parameters $a_i, b_i, z_i, m_\nu, m$ and $\mu$. $a_i$ and $z_i$ are determined up to the parameters $m_\nu$ and $t$ in Eqs.~\eqref{} and \eqref{}. $m_\nu$ and the ratio $\frac{m}{\mu}$ are determined in Eq.~\eqref{param1}. \ldots  }

Next we discuss the second scenario with four right-handed neutrinos, where $\varphi_z$ is absent in the seesaw formula.

\subsection*{\large{Scenario 2: $\varphi_{\mc B} \sim \g{3}_2, \varphi_v \sim \gb{3}_1, \varphi_{\mc A} \sim \gb{3}_2$}}
In this case, $\mc W=v$ and $\mu=m$. With the form of $\mc W$ given by \eqref{w100}, we have
\begin{align}
\matc{v_1\\ v_2\\ v_3} &= \sqrt{m m_v'}\matc{2\\ -1\\ e^{i\delta}}, \label{vvev}
\end{align}
where $m_v'$ is another mass parameter. Unlike the $\varphi_z$ of scenario 1, the vacuum alignment of $\varphi_v$ here does not depend on $\vev{\varphi_\mc B}$.

The second term in Eq.~\eqref{mastereq} becomes
\begin{align}
\mc S_2 =\frac{1}{m}vv^t= 6 m_v'\ \mc U_{TBM} (\delta)\ \text{diag}(1,0,0)\ \mc U_{TBM}^t (\delta). \label{S2'}
\end{align}
Combining with $\mc S_1$, we express the light neutrino masses in terms of the parameters  $m_{\nu}$ and $m_v'$:
\begin{align}
m_{\nu_1} =  -m_{\nu} + 6m_v',\quad m_{\nu_2} = \frac{1}{2}m_{\nu},\quad m_{\nu_3} = -m_{\nu},
\end{align}
yielding the same mass spectrum as in Eq.~\eqref{numass2}.
Using oscillation data for normal ordering \cite{tanabashi2018review} and the circle diagram in Fig.~\ref{fig:numass}, the parameters are 
\begin{align}
|m_\nu| = 57.8\ \text{meV},\quad |m_v'| = 5.03\ \text{meV}\ \text{or}\ 14.2\ \text{meV}. \label{paramv}
\end{align}

The mass parameters we have introduced so far are either completely determined from oscillation data or depend only on $b_1$, $b_2$, $b_3$ and $m$. Hence, there are only four undetermined parameters.

\subsection{TBM Mixing and the Familon Vacuum Structure}
Central to the TBM seesaw mixing are Eqs.~\eqref{avev} and \eqref{zvev}, which align the familons $\varphi_{\mc A}$ and $\varphi_z$  to $\varphi_{\mc B}$ in vacuum. Suggestively, $\varphi_{\mc A}$ and $\varphi_z$ can be expressed as quadratic functions of $\varphi_{\mc B}$ in vacuum:
\begin{align}
\begin{aligned}
    \vev{\varphi_{\mc A}} &= \sqrt{\frac{m_\nu}{b_1 b_2 b_3}} \left(
			\begin{array}{ccc}
 			0 & -e^{i\delta} & 0 \\
 			1 & 0 & 0 \\
 			0 & 0 & 1 \\
			\end{array}
		\right) \vev{\left(\varphi_{\mc B} \cdot \varphi_{\mc B}\right)_{\gb{3}_2}}, \\
    \vev{\varphi_z} &= \frac{m_{bz}^2}{b_1 b_2 b_3} \left(
			\begin{array}{ccc}
 			1 & 0 & 0 \\
 			0 & 0 & -2 \\
 			0 & 1 & 0 \\
			\end{array}
		\right) \vev{\left(\varphi_{\mc B} \cdot \varphi_{\mc B}\right)_{\gb{3}_2}}.
\end{aligned}\label{alignment}
\end{align}
Eq.~\eqref{alignment} is expected to come from the minimization of the familon potential in vacuum.

If we assume a simple vacuum alignment for $\varphi_{\mc B}$, setting $\vev{\varphi_{\mc B}} \sim b\ (1,1,1)^t$, thus reducing the number of undetermined mass parameters to two, $\varphi_{\mc A}$ and $\varphi_z$ in scenario 1 are also aligned in ``crystallographic'' directions:
\begin{align}
    \matc{a_1\\ a_2\\ a_3} &= \sqrt{m_\nu b} \matc{-e^{i\delta}\\ 1\\ 1}, \quad \matc{z_1\\ z_2\\ z_3} = \frac{m_{bz}^2}{b}\matc{1\\ -2\\ 1}. 
\end{align}
In scenario 2, $\varphi_z$ has similar vacuum alignment independent of $\vev{\varphi_{\mc B}}$.

\blu{In the next subsection we calculate the right-handed neutrino masses from diagonalization of the Majorana matrix. As we will see, setting $b_1 = b_2 = b_3 \equiv b$ greatly simplifies the analysis and yields interesting cases of degeneracy in the mass spectrum.}

\subsection{Right-handed Neutrino Mass Spectrum}
\blu{We now explore the right-handed neutrino masses in the two scenarios discussed before. Although these scenarios yield identical light neutrino mass spectrum, their predictions for the right-handed neutrinos are quite different. }

\subsection*{\large{Scenario 1: $\varphi_{\mc B} \sim \g{3}_2, \varphi_z \sim \gb{3}_2, \varphi_{\mc A} \sim \gb{3}_2$}}
\begin{figure}[!h]
    \centering
    \includegraphics[width=0.78\textwidth]{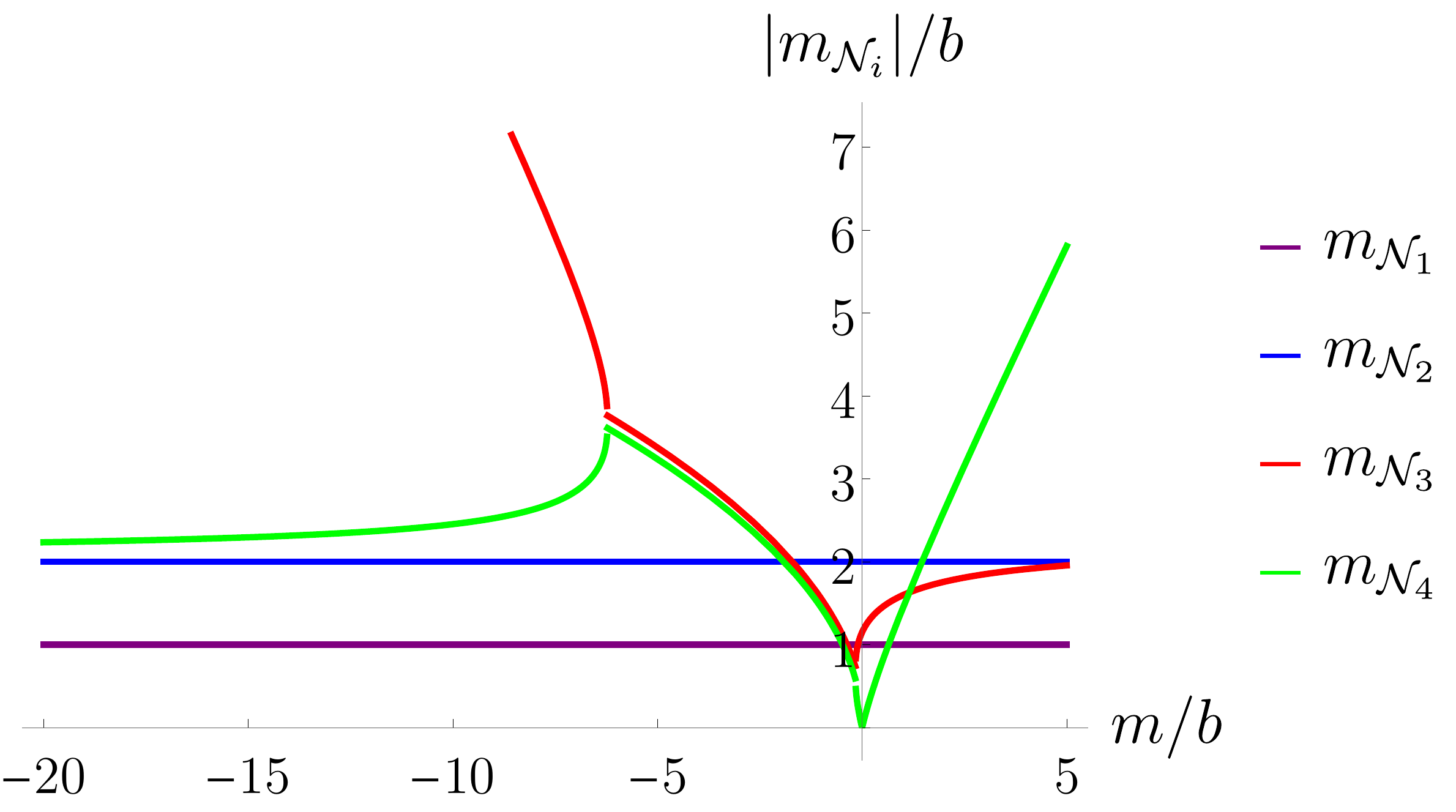}
    \caption{Normalized right-handed neutrino mass spectrum. A small number has been added to $m_{\mc N_3}$ to separate it from $m_{\mc N_4}$ in the degenerate region  $-6.23\leqslant \frac{m}{b} \leqslant -0.16$. Note that these masses become complex in this region, thus their modulus is degenerate, although they have a relative phase.}
    \label{fig:my_label1}
\end{figure}

In this case, the Majorana matrix is
\begin{align}
    \mathcal{M} &= \left(
\begin{array}{cccc}
 0 & b_2 & b_3 & \dfrac{m_{bz}^2}{b_1} \\[0.3em]
 b_2 & 0 & b_1 & -\dfrac{2 m_{bz}^2}{b_3} \\[0.3em]
 b_3 & b_1 & 0 &  \dfrac{m_{bz}^2}{b_2}\\[0.3em]
 \dfrac{m_{bz}^2}{b_1} & -\dfrac{2 m_{bz}^2}{b_3} & \dfrac{m_{bz}^2}{b_2} & m \\
\end{array}
\right) \label{Majorana}
\end{align}
where $b_i \neq 0$. From Eqs.~\eqref{muval} and \eqref{param1},
\begin{align}
    &\dfrac{b_1 b_2 b_3 m}{b_1 b_2 b_3 m + 6m_{bz}^4} = 0.48 \equiv \dfrac{1}{k} \\
    \implies &m_{bz}^4 = \dfrac{k-1}{6} m b_1 b_2 b_3.
\end{align}
Setting $b_1=b_2=b_3\equiv b$, the characteristic equation for $\mathcal{M}$ becomes
\begin{align}
    x^4 - mx^3 - b(3b+m(k-1))x^2 + b^2 \left(m(k+2)-2b \right)x +2b^3 km = 0.
\end{align}
Its solutions yield the four right-handed neutrino masses: 
\begin{align}
\begin{split}
    m_{\mc N_1} &= -b,\\
    m_{\mc N_2} &= 2b, \\
    m_{\mc N_3} &= \frac{b}{2} \left( \left(\frac{m}{b}-1\right) - \sqrt{\left(\frac{m}{b}-1\right)^2+4k\frac{m}{b}} \right), \\
    m_{\mc N_4} &= \frac{b}{2} \left( \left(\frac{m}{b}-1\right) + \sqrt{\left(\frac{m}{b}-1\right)^2+4k\frac{m}{b}} \right).
\end{split}
\end{align}
In Fig.~\ref{fig:my_label1}, we plot the normalized mass spectrum with respect to $\frac{m}{b}$.

There are several interesting cases of degeneracy in the mass spectrum. $m_{\mc N_3}$ and $m_{\mc N_4}$ are degenerate for $-6.23\leqslant \frac{m}{b} \leqslant -0.16$. We also have two degenerate masses at $\frac{m}{b}=0.65,\ 1,\ 1.46$. And three of the masses are degenerate for $\frac{m}{b}=-1.91,\ -0.48$. Such degeneracies or near degeneracies in the right-handed neutrino spectrum may be interesting from the point of leptogenesis, where an enhancement of the $CP$ asymmetry is achieved for near-degenerate masses \cite{Pilaftsis:2003gt, *Dev:2017wwc, *Pilaftsis:2009pk}.

\subsection*{\large{Scenario 2: $\varphi_{\mc B} \sim \g{3}_2, \varphi_v \sim \gb{3}_1, \varphi_{\mc A} \sim \gb{3}_2$}}
In this case the Majorana matrix $\mc M$ is simpler:
\begin{align}
\mc M = \left(
\begin{array}{cccc}
 0 & b_2 & b_3 & 0 \\
 b_2 & 0 & b_1 & 0 \\
 b_3 & b_1 & 0 &  0\\
 0 & 0 & 0 & m \\
\end{array}
\right). \label{Majorana2}
\end{align}
Again setting $b_1=b_2=b_3\equiv b$, the right-handed neutrino masses are given by the following eigenvalues of $\mc M$:
\begin{equation}
m_{\mc N_1} = -b,\quad
m_{\mc N_2} = -b,\quad
m_{\mc N_3} = 2b,\quad
m_{\mc N_4} = m.
\end{equation}
Unlike the previous scenario, the masses are dependent on either $b$ or $m$, but not both. The first two masses are degenerate. 

This ends our discussion of neutrino masses and mixings. In the next section, we calculate the $\cancel{CP}$ phases predicted by the asymmetric texture and discuss their implication for neutrinoless double-beta decay.

\section{$\cancel{CP}$ Phases and $\left|m_{\beta\beta} \right|$}\label{sec:4}
In order to analyze the $\cancel{CP}$ phases in the asymmetric texture, consider the Jarlskog-Greenberg invariant $\mc J$ \cite{jarlskog1, *greenberg1} given by
\begin{equation}\label{eq:jarlkoginvgen}
\mathrm{Im} (\mc U_{ij}\ \mc U_{kl}\ \mc U_{il}^*\ \mc U_{kj}^*)=\mathcal{J}\sum_{n,m}\epsilon_{ikm}\ \epsilon_{jln}.
\end{equation}
Letting $i=j=1$ and $k=l=2$ in the above equation ``fixes'' the sign so that%\footnote{Note: There is still an ambiguity on the sign of $\mathcal{J}$.}
\begin{equation}\label{eq:jarlskoginv}
\mathcal{J} = \mathrm{Im} (\mc U_{11}\ \mc U_{22}\ \mc U_{12}^*\ \mc U_{21}^*).
\end{equation}
The two analogous invariants associated with the Majorana phases \cite{jenkins2008rephasing} are then
\begin{equation}\label{eq:MajInvs}
\begin{aligned}
\mathcal{I}_1 &= \mathrm{Im}(\mc U_{12}\ \mc U_{11}^*)^2, ~~~~ \mathcal{I}_2 = \mathrm{Im}(\mc U_{13}\ \mc U_{11}^*)^2.
\end{aligned}
\end{equation}
Next, let $\mc U = \mc U_{PMNS}$ in the PDG convention \cite{tanabashi2018review}, so that
\begin{equation}\label{eq:mnsppdg}
 \mc U=\left(
\begin{array}{ccc}
 c_{12} c_{13} & c_{13} s_{12} & e^{-i \delta_{CP} } s_{13} \\
 -c_{23} s_{12}-c_{12} s_{13} s_{23}e^{i \delta_{CP} }  & c_{12} c_{23}- s_{12} s_{13} s_{23}e^{i \delta_{CP} } & c_{13} s_{23} \\
 s_{12} s_{23}-c_{12} c_{23}  s_{13}e^{i \delta_{CP} } & -c_{12} s_{23}-c_{23}  s_{12} s_{13}e^{i \delta_{CP} } & c_{13} c_{23} \\
\end{array}
\right)P\\ 
\end{equation}
where $P={\rm diag} \left(1,e^\frac{i\alpha_{21}}{2},e^\frac{i\alpha_{31}}{2}\right)$ is a diagonal matrix of Majorana phases,  $s_{ij}=\sin{\theta_{ij}}$ and $c_{ij}=\cos{\theta_{ij}}$.  The Jarlskog-Greenberg invariant from Eq.~\eqref{eq:jarlskoginv} in the PDG convention is given as
\begin{equation}\label{eq:jarlskogpdg}
\mathcal{J}_{PDG}=c_{12} c_{13}^2 c_{23} s_{12} s_{13} s_{23} \sin {\delta_{CP} }=  \frac{1}{8}  s'_{12} s'_{13} s'_{23}  c_{13} \sin{\delta_{CP}},
\end{equation}
where $s'_{ij}= \sin{2\theta_{ij}}$.  Finally, the PDG Majorana invariants  are given by
\begin{equation}\label{eq:MajInvspdg}
\begin{aligned}
\mathcal{I}_1^{PDG}&= c_{12}^2 c_{13}^4 s_{12}^2 \sin {\alpha_{21}},~~~
\mathcal{I}_2^{PDG} = c_{12}^2 c_{13}^2 s_{13}^2 \sin {(\alpha_{31}-2 \delta_{CP}) }.
\end{aligned}
\end{equation}
From Eqs.~\eqref{eq:jarlskogpdg} and \eqref{eq:MajInvspdg}, it is possible to extract the three $\cancel{CP}$ phases knowing the values of the angles in the PDG convention.

The PMNS mixing matrix resulting from the asymmetric texture \cite{rrx2018asymmetric} is parametrized as $\mc U_{PMNS} = {\mc U^{(-1)}}^\dagger \ \mc U_{TBM}(\delta)$, where
\begin{equation}\label{eq:ul}
\mc U^{(-1)} =\left(
\begin{array}{ccc}
 1-\left(\frac{2}{9 A^2}+\frac{1}{18}\right) \lambda ^2 & \frac{\lambda }{3} & \frac{2 \lambda }{3 A} \\[0.5em]
 -\frac{\lambda }{3} & 1-\frac{\lambda ^2}{18} & A \lambda ^2 \\[0.5em]
 -\frac{2 \lambda }{3 A} & \left(-A-\frac{2}{9 A}\right) \lambda ^2 & 1-\frac{2 \lambda ^2}{9 A^2} \\
\end{array}
\right)+\mathcal{O}(\lambda^3)
\end{equation}
From  $\mc U_{PMNS}$, we calculate the mixing angles in the PDG convention, cf.~Eq.~\eqref{eq:mnsppdg} as
\begin{equation}\label{eq:anglesasym}
\begin{aligned}
\theta_{13}&=\frac{\lambda  \sqrt{A^2+4 A \cos {\delta} +4}}{3 \sqrt{2} A}+\mathcal{O}(\lambda^3),\\[1em]
\theta_{23}&=\frac{\pi }{4}+\frac{ (4-4 (9 A^3+A) \cos {\delta }-A^2)}{36 A^2}\lambda ^2+\mathcal{O}(\lambda^3),\\[1em]
\theta_{12}&=\sin ^{-1}\left(\frac{1}{\sqrt{3}}\right)+\frac{2 \cos {\delta} -A}{3 \sqrt{2} A}\lambda +\frac{\sin^2{\delta }}{9 \sqrt{2} A^2}\lambda^2+\mathcal{O}(\lambda^3).
\end{aligned}
\end{equation}
Notice that angles in the above equation are just perturbative corrections in the expansion parameter $\lambda$ to the initial angle starting points of $ \mc U_{TBM}(\delta=0)$.

Using the perturbatively calculated angles of Eq.~\eqref{eq:anglesasym}, it is possible to find the Jarlskog-Greenberg invariant of Eq.~\eqref{eq:jarlskogpdg} and Majorana invariants of Eq.~\eqref{eq:MajInvspdg}:
\begin{equation}\label{eq:jarlskogpert}
\begin{aligned}
\mathcal{J} &= \frac{\lambda \sin {\delta }}{9 A}-\frac{\lambda ^2 \sin {\delta }}{27 A} + \mc O(\lambda^3),\\
\mc I_1 &= \frac{4 \lambda  \sin {\delta }}{9 A}-\frac{2 \lambda ^2 \sin {\delta }\ (A-2 \cos {\delta})}{27 A^2} + \mc O(\lambda^3), \\
\mc I_2 &= \frac{4 \lambda ^2 \sin {\delta }\ (A+2 \cos {\delta })}{27 A^2} + \mc O(\lambda^3).
\end{aligned}
\end{equation}
Note that in the asymmetric texture, all the invariants have the same sign, determined by  $\sin(\delta)$.

Following the results of Ref.~\cite{rrx2018asymmetric}, we calculate, to $\mc O(\lambda^3)$, the mixing angles as
\begin{equation}\label{eq:anglesasymnum}
\begin{aligned}
\theta_{13}&= 8.33^{\circ}, ~~\theta_{23}=44.87^{\circ},~~\theta_{12}=34.09^{\circ},
\end{aligned}
\end{equation}
and the invariants as 
\begin{equation}\label{eq:jarlkoginv}
\begin{aligned}
\mathcal{J} &= 0.028, \qquad \qquad \mathcal{J} = -0.028,\\
\mathcal{I}_1 &= 0.106, ~~\quad \mathrm{or} \quad~ \mathcal{I}_1 = -0.106,\\
\mathcal{I}_2 &= 0.011, \qquad \qquad \mathcal{I}_2 = -0.011.
\end{aligned}
\end{equation}
The above values can be used to extract values for the $\cancel{CP}$ phases [cf. Eqs.~\eqref{eq:jarlskogpdg} and\eqref{eq:MajInvspdg}]:
\begin{equation}\label{eq:sinphases}
\begin{aligned}
\sin \delta_{CP} &= 0.854,\qquad \qquad \quad \qquad \sin \delta_{CP} = -0.854,\\
\sin \alpha_{21} &= 0.515,~~~\quad  \mathrm{or} \quad \qquad \quad~ \sin \alpha_{21} = -0.515,\\
\sin(\alpha_{31}-2\delta_{CP})&= 0.809, \quad \qquad \quad  \sin(\alpha_{31}-2\delta_{CP})= -0.809.
\end{aligned}
\end{equation}

\noindent With the three light neutrino masses and the Dirac and Majorana phases determined, we can now express the effective Majorana mass parameter in neutrinoless double-beta decay as \cite{bilenky2015neutrinoless}
\begin{align}
    | m_{\beta \beta}  | &= \left| c_{13}^2 c^2_{12} m_{\nu_1} + c_{13}^2 s^2_{12} e^{i\alpha_{21}} m_{\nu_2} + s_{13}^2 m_{\nu_3} e^{i(\alpha_{31}-2\delta_{CP})} \right| \label{mbb}
\end{align}
Note that in Eq.~\eqref{eq:sinphases} all the signs are either positive or negative. This does not make any difference in evaluating $\left|m_{\beta \beta}\right|$ in  Eq.~\eqref{mbb}. However, there are ambiguities in the signs of the light neutrino masses. For example, in Eq.~\eqref{numass3}, these masses have been expressed in terms of $\frac{m}{\mu}$ and $m_{\nu}$. The absolute value of $\frac{m}{\mu}$ and $m_\nu$ has been determined in Eq.~\eqref{param1}, but the signs remain undetermined. Depending on which sign is realized, $|m_{\beta \beta}|$ is predicted to be one of the following:
\begin{align}
| m_{\beta \beta}  | &= 13.02\ \quad \text{or}\quad 25.21\ \text{meV}.
\end{align}

The most stringent experimental upper bound on $|m_{\beta \beta}|$ is in between $61$ and $165$ $\text{meV}$, reported recently by the KamLAND-Zen experiment \cite{gando2016search}.\footnote{See \cite{Anton:2019wmi, *Agostini:2018tnm, *Cappuzzello:2018wek, *Alenkov:2019jis} for other recent results.} Both of our  predicted values are within an order of magnitude of this limit. 

This ends our discussion of the seesaw sector. In the next section, we summarize the components and predictions of the model.

\section{Summary of the Model}\label{sec:6}
\blu{We proposed a phenomenologically successful framework --- a diagonal $Y^{(\frac{2}{3})}$, asymmetric $Y^{(-\frac{1}{3})}$ and $Y^{(-1)}$ related by $SU(5)$ grand unification, and a complex-TBM seesaw mixing --- in Ref.~\cite{rrx2018asymmetric}. In Ref.~\cite{Perez:2019aqq}, we built a model based on $SU(5) \times \mc T_{13}$ symmetry that constructs the asymmetric $Y^{(-\frac{1}{3})}$ and $Y^{(-1)}$ textures. In this paper, we show how the diagonal $Y^{(\frac{2}{3})}$ texture and the complex-TBM seesaw mixing follows from the $SU(5) \times \mc T_{13}$ symmetry. We now put all the pieces of the puzzle together to construct a unified model that describes both quarks and leptons.}

\blu{The gauge and family symmetry of the model are $SU(5)$ and $\mc T_{13}$, respectively. This still allows some unwanted operators at the tree level. In Appendix~\ref{app:shape}, we show that such operators can be prevented by introducing a $\mc Z_n$ ``shaping'' symmetry, where $n$ is determined to be $14$ for the scenario with no $\varphi_v$, and $12$ with no $\varphi_z$. Thus the full symmetry of the unified model is $SU(5) \times \mc T_{13} \times \mc Z_n$.}

\subsection{Particle Content and their Transformation Properties}
The tree-level Lagrangian of the model is
\begin{align} \label{lagrangian}
\mc L &= y_0 T \overline{\Delta} H_{\gb{5}} + y_1 F \Delta\varphi^{(1)} + y_2 F \Delta\varphi^{(2)} + y_3 \overline{\Delta} \Delta \varphi^{(3)} + y_4 F \Delta\varphi^{(4)}  \nonumber \\
 &+ y_5 F \Delta\varphi^{(5)} + M_\Delta \overline{\Delta} \Delta + y_6 F\overline{\Sigma} H_{\overline{\g{45}}}+ y_7 T\Sigma {\varphi^{(6)}}+M_\Sigma \overline{\Sigma} \Sigma \nonumber \\
 &+ y_8 T \Gamma \varphi_{\g{3}_1}^{(t)} + y_9 T \Omega \varphi_{\g{3}_1}^{(t)} + y_{10} T \Theta \varphi_{\g{3}_1}^{(t)} + y_{11} T \overline{\Gamma} \bar{H}_{\g{5}} + y_{12} \Gamma \overline{\Omega} \varphi_{\g{3}_2} \nonumber \\
 &+ y_{13} \Theta \overline{\Theta} \varphi_{\g{3}_2} + y_{14} \Gamma \overline{\Theta} \varphi_{\g{3}_2} + M_\Gamma \overline{\Gamma} \Gamma + M_\Omega \overline{\Omega} \Omega + M_\Theta \overline{\Theta} \Theta \nonumber \\
 &+ y_{\mc A} F \Lambda \bar{H}_{\g{5}} + y'_{\mc A} \bar{N} \overline{\Lambda} \varphi_{\mc A} + y_{\mc B} \bar{N} \bar{N} \varphi_{\mc B} + y'_v\bar{N} \overline{\Lambda} \varphi_v  + M_\Lambda \overline{\Lambda} \Lambda \nonumber \\
 &+  y_z \bar{N} \bar{N}_4 \varphi_z + m \bar{N}_4 \bar{N}_4,
 \end{align}
 where only one of $\varphi_z$ and $\varphi_v$ is present. The first two lines describe the down-type quarks and charged leptons, the next two yield the up-type quark masses and the last two depict the seesaw sector of the model. The $\mc Z_{n}$ symmetry ensures that the familons and messengers in one sector do not mix with fields in the other sector. In Table \ref{table:summary}, we show the transformation properties of the fields in each sector.

\begin{table}[ht]\centering
\renewcommand\arraystretch{1.25}
\begin{tabularx}{\textwidth}{@{}l | Y Y | Y Y Y Y Y Y Y Y Y Y @{}}
\toprule
& \multicolumn{2}{c | }{Higgs} & \multicolumn{10}{c }{Down-type quark and Charged-lepton Sector} \\
\hline
     Fields & $H_{\gb{5}}$ & $H_{\overline{\g{45}}}$ & $F$ & $T$ &  $\Delta$  & $\Sigma$ & $\varphi^{(1)}$ & $\varphi^{(2)}$ & $\varphi^{(3)}$ & $\varphi^{(4)}$ & $\varphi^{(5)}$&${\varphi^{(6)}}$\\ 
\hline
$SU(5)$    & $\overline{ \g{5}}$&$\overline{\g{45}}$ & $\overline{ \g{5}}$ & $\g{10}$ & $\g{5}$ & $\overline{ \g{10}}$ & $\g{1}$ & $\g{1}$ & $\g{1}$ & $\g{1}$ & $\g{1}$ & $\g{1}$ \\ 

$\mc T_{13}$ & $\g{1}$ & $\g{1}$ & $\g{3}_2$  &  $\g{3}_1$ & $\g{3}_2$ & $\g{3}_1$ & $\gb{3}_2$ & $\g{3}_2$ & $\gb{3}_1$  & $\gb{3}_2$ & $\gb{3}_1$ & $\g{3}_2$\\ 

$\mathcal{Z}_{14}$    & $\g{\eta^3}$ & $\g{\eta^4}$ & $\g{\eta^1}$& $\g{\eta^1}$ & $\g{\eta^4}$  & $\g{\eta^5}$ & $\g{\eta^9}$   & $\g{\eta^9}$ & $\g{1}$  & $\g{\eta^9}$ & $\g{\eta^9}$ & $\g{\eta^8}$ \\ %[3 4 1 1 4 5 9 9 0 9 9 8]	
$\mathcal{Z}_{12}$    & $\g{\zeta^3}$ & $\g{\zeta^1}$ & $\g{\zeta^1}$& $\g{1}$ & $\g{\zeta^3}$  & $\g{\zeta^2}$ & $\g{\zeta^{8}}$   & $\g{\zeta^{8}}$ & $\g{1}$  & $\g{\zeta^{8}}$ & $\g{\zeta^{8}}$ & $\g{\zeta^{10}}$ \\ %[ 3  1  1  0  3  2  8  8  0  8  8 10]

\end{tabularx}
\vspace{0.3cm}
\begin{tabularx}{\textwidth}{@{}l | Y Y Y Y Y | Y Y Y Y Y Y Y @{}}
\toprule
 & \multicolumn{5}{c |}{Up-type quark Sector} & \multicolumn{6}{c}{seesaw Sector}\\
\hline
     Fields & $\Gamma$ & $\Omega$ & $\Theta$ & $\varphi_{\g{3}_1}^{(t)}$ & $\varphi_{\g{3}_2}$  & $\bar{N}$ & $\bar{N}_4$ & $\Lambda$ & $\varphi_{\mc A}$ & $\varphi_{\mc B}$ & $\varphi_{z}$ & $\varphi_{v}$ \\ 
\hline
$SU(5)$    & $\overline{ \g{10}}$ & $\overline{\g{10}}$ & $\overline{ \g{10}}$ & $\g{1}$ & $\g{1}$ & $\g{1}$ & $\g{1}$ & $\g{1}$ & $\g{1}$ & $\g{1}$ & $\g{1}$ & $\g{1}$ \\ 

$\mc T_{13}$ &  $\g{3}_2$ & $\gb{3}_2$ & $\gb{3}_1$ & $\g{3}_1$ & $\g{3}_2$  & $\g{3}_2$ & $\g{1}$ & $\gb{3}_1$ & $\gb{3}_2$ & $\g{3}_2$  & $\gb{3}_2$ & $\gb{3}_1$ \\ 
$\mathcal{Z}_{14}$    & $\g{\eta^{12}}$ & $\g{\eta^{12}}$ & $\g{\eta^{12}}$& $\g{\eta^1}$ & $\g{1}$  & $\g{\eta^5}$ & $\g{\eta^7}$ & $\g{\eta^2}$ & $\g{\eta^{11}}$ & $\g{\eta^4}$  & $\g{\eta^2}$  & $\times$  \\
$\mathcal{Z}_{12}$    & $\g{\zeta^{9}}$ & $\g{\zeta^{9}}$ & $\g{\zeta^{9}}$& $\g{\zeta^3}$ & $\g{1}$  & $\g{\zeta^3}$ & $\g{1}$ & $\g{\zeta^2}$ & $\g{\zeta^{11}}$ & $\g{\zeta^6}$  & $\times$  &  $\g{\zeta^2}$ \\ %[ 9  9  9  3  0  3  0  2 11  6  9  2]
\botrule 	
\end{tabularx} 
\caption{Charge assignments of matter, Higgs, messenger and familon fields. $\mc Z_{14}$ charges apply for the scenario with no $\varphi_v$ and $\mc Z_{12}$ for no $\varphi_z$. The symbol $\times$ implies `not applicable'. Here $\eta^{14} = \zeta^{12} = 1$.}
\label{table:summary}
\end{table}

\subsection{Familon Vacuum Structure}
The familons in the quark and charged-lepton sectors have a ``crystallographic'' feature in vacuum, in the sense that they are aligned along sides or face diagonals of a cube. The seesaw sector familons, which depend on $b_1, b_2, b_3$, are also similarly aligned if we set $b_1 = b_2 = b_3 \equiv b$. In Table~\ref{table:vev}, we list all vacuum alignments.
\begin{table}[!ht] 
\renewcommand*{\arraystretch}{1.5}
\begin{tabularx}{\textwidth}{Y Y}
\toprule
Down-type quark and charged-lepton sector & Up-type quark and seesaw sector\\
\hline
     $\begin{aligned}
         \vev{\varphi^{(1)}} &\sim m_b(1,0,0) \\[0.4em]
         \vev{\varphi^{(2)}} &\sim d\lambda~ m_b(0,1,0)\\[0.4em]
         \vev{\varphi^{(3)}} &\sim b\lambda^3~ m_b(0,0,1)\\[0.4em]
         \vev{\varphi^{(4)}} &\sim a\lambda^3~ m_b (0,1,1)\\[0.4em]
         \vev{\varphi^{(5)}} &\sim g\lambda^2~ m_b(1,0,1)\\[0.4em]
         \vev{\varphi^{(6)}} &\sim c\lambda^2~ m_b(0,0,1)
     \end{aligned}$  & 
     $\begin{aligned}
         \vev{\varphi_{\g{3}_1}^{(t)}} &\sim m_t (1,0,0)\\[0.3em]
         \vev{\varphi_{\g{3}_2}} &\sim \lambda^4 (1,0,0)\\[0.3em]
         \vev{\varphi_{\mc B}} &\sim  b(1, 1, 1)\\[0.3em]
         \vev{\varphi_{\mc A}} &\sim \sqrt{m_\nu b} (-e^{i\delta},1,1)\\[0.3em]
         \vev{\varphi_{z}} &\sim  \frac{m_{bz}^2}{b} (1, -2, 1)\\[0.3em]
         \vev{\varphi_{v}} &\sim  \sqrt{m m_2'} (2, -1, e^{i\delta})
     \end{aligned}$\\
\botrule
\end{tabularx}
\caption{Vacuum alignment of familons, setting $b_1 = b_2 = b_3 \equiv b$.}
\label{table:vev}
\end{table}

\begin{comment}
$\vev{\varphi^{(1)}} \sim m_b(1,0,0)$, & \qquad $\vev{\varphi_{\g{3}_1}^{(t)}} \sim m_t (1,0,0)$,\\
$\vev{\varphi^{(2)}} \sim d\lambda~ m_b(0,1,0)$, & \qquad $\vev{\varphi_{\g{3}_2}} \sim m_t \lambda^4 (1,0,0)$,\\
$\vev{\varphi^{(3)}} \sim b\lambda^3~ m_b(0,0,1)$, & \qquad $\vev{\varphi_{\mc B}} \sim  (b_1, b_2, b_3)$,\\
$\vev{\varphi^{(4)}} \sim a\lambda^3~ m_b (0,1,1)$, & \qquad $\vev{\varphi_{\mc A}} \sim \sqrt{m_\nu b_1 b_2 b_3} (-b_2^{-1}e^{i\delta},b_1^{-1},b_3^{-1})$,\\
$\vev{\varphi^{(5)}} \sim g\lambda^2~ m_b(1,0,1)$, & \qquad
$\vev{\varphi_{z}} \sim  m_{bz}^2 (b_1^{-1}, -2b_3^{-1}, b_2^{-1})$,\\
$\vev{\varphi^{(6)}} \sim c\lambda^2~ m_b(0,0,1)$, & \qquad $\vev{\varphi_{v}} \sim  \sqrt{m m_2'} (2, -1, e^{i\delta})$.
\end{comment}

Note that the vacuum expectation values of $\varphi_{\mathcal{A}}$ and $\varphi_v$ contain a nontrivial phase $\delta$, as required by the alignment conditions of the seesaw sector. We view this as an interesting constraint on the parameters of the familon vacuum potential, to be studied in a follow-up work.

\subsection{Predictions}
The model successfully reproduces the CKM mixing angles, Gatto relation, GUT-scale mass ratios of up-type quarks, down-type quarks and charged leptons as well as the PMNS mixing angles. 

The key predictions of the model are
\begin{itemize}
    \item leptonic $CP$ violation, with the Jarlskog-Greenberg invariant $|\mc J| = 0.028$, Majorana invariants $|\mc I_1| = 0.106$ and $|\mc I_2| = 0.011$,
    \item normal ordering of light neutrino masses: $m_{\nu_1} = 27.6$ $\text{meV}$, $m_{\nu_2}=28.9$ $\text{meV}$ and $m_{\nu_3} = 57.8$ $\text{meV}$, and
    \item invariant mass parameter in neutrinoless double-beta decay $|m_{\beta \beta}| = 13.02$  or $25.21$ $\text{meV}$.
\end{itemize}

\myred{The first prediction ($|\mc J|$) is consistent with the current PDG fit \cite{tanabashi2018review} and translates into $\delta_{CP} = \pm 0.68\pi$ \cite{rrx2018asymmetric}. Although current expected error in global fit for $\delta_{CP}$ is too wide, it is expected that next-generation experiments like DUNE \cite{Abi:2018dnh} and Hyper-K \cite{Abe:2018uyc} will measure this with $5\sigma$ precision in the next decade.}  

\myred{The second prediction for ordering of light neutrino masses can, in principle, be tested experimentally in three ways \cite{deSalas:2018bym}: (i) oscillation experiments that directly measure the sign of $\Delta m_{31}^2$, (ii) cosmological bounds on $\sum_i |m_{\nu_i}|$, and (iii) measurement of $|m_{\beta \beta}|$ in neutrinoless double-beta decay experiments. If $\sum_i |m_{\nu_i}| < 10\ \text{meV}$ or $|m_{\beta \beta}| < 10\ \text{meV}$ we can rule out inverted ordering \cite{deSalas:2018bym}, assuming neutrinos are Majorana particles; but neither of these materializes in this model. Hence, we must rely on oscillation experiments to determine the mass ordering. The current fit from various experiments (e.g. Super-Kamiokande \cite{Abe:2017aap}, T2K \cite{T2K2017}, NOvA \cite{nova2018}) gives above $3\sigma$ preference for normal over inverted ordering. A $3\sigma$ rejection of the wrong mass ordering will be obtained in Hyper-K \cite{Abe:2018uyc} after five years of data taking. DUNE will be able to measure the mass ordering with a significance above $5\sigma$ after $7$ years of data taking \cite{Abi:2018dnh}. }

\myred{The second prediction also gives $\sum_i |m_{\nu_i}| = 114.3$ $\text{meV}$, to be compared with the strictest cosmological upper bound of $120$ $\text{meV}$ reported recently by combining various sources of data by the Planck collaboration \cite{aghanim2018planck}. Combining the data from large scale structure surveys, e.g., Euclid \cite{Amendola:2016saw} and LSST \cite{Abell:2009aa} to DESI \cite{Levi:2013gra, *Aghamousa:2016zmz}, and WFIRST \cite{Spergel:2015sza},  the error margin on $\sum_i |m_{\nu_i}|$ will be constrained to less than $11$ $\text{meV}$ \cite{Font-Ribera:2013rwa}, and $8$ $\text{meV}$ \cite{Jain:2015cpa}, respectively. These estimates can test our prediction in coming years.}

\myred{The third prediction is consistent with the recently reported upper bound of $61$--$165$ $\text{meV}$ by the KamLAND-Zen experiment \cite{gando2016search} and is expected to be tested in next-generation experiments in R\&D \cite{Barabash:2019drl} (LEGEND: $11$--$28$ $\text{meV}$ \cite{Abgrall:2017syy}, CUPID: $6$--$17$ $\text{meV}$ \cite{Wang:2015raa, *Wang:2015taa}, nEXO: $8$--$22$ $\text{meV}$ \cite{Albert:2017hjq}, SNO+-II: $20$ - $70$ $\text{meV}$ \cite{Andringa:2015tza, *Fischer:2018squ}, AMoRE-II: $15$--$30$ $\text{meV}$ \cite{Jo:2017jod}, PandaX-III: $20$--$55$ $\text{meV}$ \cite{Chen:2016qcd}), which will be sensitive to the range of our predictions. If either of our predictions is correct, these experiments will detect neutrinoless double-beta decay \cite{Barabash:2019drl}.}

\section{Theoretical Musings}\label{sec:7}
We have presented an asymmetric unified texture of quarks and leptons. Under the grand-unified  $SU(5)$ times the discrete family symmetry $ \mc T_{13}$ its fermion content $T, F, \bar{N}, \bar{N}_4$ is,
$$
(\bf 10,\bf \g{3}_2)\oplus (\bf\bar 5,\bf \g{3}_1) \oplus (\bf 1,\bf \g{3}_2) \oplus (\bf 1, \bf 1).
$$
%which includes four heavy Majorana neutrinos. The reactor angle in the PMNS lepton mixing matrix comes from the diagonalization of the charged-lepton Yukawa couplings while the two large atmospheric and solar angles come from the seesaw structure.

%The 39-element Frobenius group $\mc T_{13}$ has three complex representations, denoted by $\bf \g{3}_1$, $\bf \g{3}_2$, $\bf 1'$ and their conjugates. 

By upgrading $SU(5)$ to $SO(10)$, we get a simpler particle content 
\be
SO(10) \times\mc T_{13}:~~~~~ (\bf 16,\bf \g{3}_2) \oplus (\bf 10,\bf \g{3}_1) \oplus (\bf 1, \bf 1). \label{SO(10)}
\ee
The decomposition $SO(10) \supset SU(5)\times U(1)$
$$
\bf 16=\bf 10_{-1} \oplus \bf\bar 5_{3} \oplus \bf 1_{-5},\qquad \bf 10=\bf 5_2 \oplus \bf\bar 5_{-2}, 
$$
shows that the  $\bf 5$ in the $\bf 10$ can couple to the $\bf\bar 5$ in the $\bf 16$ and acquire a heavy $\Delta I_w=0$ vectorlike mass. This leaves $T$ and $\bar{N}$ embedded inside the $\g{16}$ and explains their similar labeling.

The grand-unified group above $SO(10)$ is the exceptional group $E_6$. Its complex 27-dimensional fundamental representation decomposes under $E_6 \supset SO(10) \times U(1)$ as 
$$\bf 27=\bf 16_1 \oplus \bf 10_{-2} \oplus 1_{4},
$$ which are precisely the representations  in the asymmetric texture's particle set. It is a suggestive pattern:  matching the representations of the gauge group to those of the discrete group. The mother symmetry could be $E_6\times \mc G_f$, where $\mc G_f$ is a \myred{continuous} group that contains $\mc T_{13}$ \cite{Rachlin:2017rvm, *Merle:2011vy, *Luhn:2011ip}.

There are several ways to see how  $\mc T_{13}$ fits in a continuous group. The first is
$$
G_2\supset  \mc P\mc S\mc L_2(13)\supset  \mc Z_{13}\rtimes \mc Z_6\supset \mc \mc Z_{13}\rtimes \mc Z_3= \mc T_{13}
$$
with the embeddings 
$$
\bf 7 \rightarrow \bf \begin{matrix} \bf 7\cr \bf 7'\end{matrix}\rightarrow\bf 6\rightarrow \begin{matrix} \bf \g{3}_1\cr \bf \g{3}_2\end{matrix}$$
which occurs through the embedding of $G_2$'s real $\bf 7$ representation. 

\noindent The second way is more direct with 
\begin{align*}
\begin{gathered}
G_2\supset SU(3)\supset \mc Z_{13}\rtimes \mc Z_3=\mc T_{13}\\
\bf 7 \rightarrow \bf \bf 1 \oplus \bf 3 \oplus \bf\bar 3\rightarrow \begin{matrix} \bf \g{3}_1\cr \bf \g{3}_2\end{matrix}.
\end{gathered}
\end{align*}
All paths seem to lead to the continuous exceptional group $G_2$ whose  seven-dimensional representation describes the coset manifold of $11$-dimensional space-time.

It would be nice  to obtain the particle content of Eq.~\eqref{SO(10)} as the result of a spontaneously broken theory. For this we need both $SO(10)$ and $\mc T_{13}$ to be extended so as to be able to pair their representations as described. To that purpose the representations must be tagged. On the gauge side it is easy since the $E_6$ decomposition into $SO(10)$ contains a continuous $U(1)$. On the family side, there is no available tag, so we must invent one.

%On the other hand,  the Abelian discrete symmetry that explains the dimension five and higher operators is $\mc Z_{12}$, which can be used to tag the three $\mc T_{13}$ representations. If their charges match, they pair up.

\section{Conclusion}\label{sec:8}
Continuing from our recent work in Ref.~\cite{Perez:2019aqq}, we have derived the up-type quark sector of the asymmetric texture \cite{rrx2018asymmetric} and the complex-tribimaximal seesaw mixing  from an $SU(5) \times \mc T_{13} \times \mc Z_n$ symmetry. This results in a unified model for quarks and leptons from $SU(5)$ gauge unification and $\mc T_{13}$ family symmetry.\footnote{See, for example, \cite{deMedeirosVarzielas:2017sdv, *Rachlin:2018xkm, *Chen:2013wba} for other unified models employing gauge and family symmetry.}

$\mc T_{13}$, an off-the-beaten-road subgroup of $SU(3)$, is a powerful family symmetry. Its ability to label each matrix element of a Yukawa texture with a distinct $\mc Z_{13}$ charge makes it an ideal candidate for constructing the asymmetric texture. Although not evident straightaway, we showed in Ref.~\cite{Perez:2019aqq} that it is capable of naturally producing the zero-subdeterminant condition of the $Y^{(-\frac{1}{3})}$ and $Y^{(-1)}$ textures. In this paper we have shown how it yields the hierarchical diagonal structure of the $Y^{(\frac{2}{3})}$ texture.

What comes as a true surprise is how the complex-tribimaximal mixing arises from the familon vacuum structure in $\mc T_{13}$. The Clebsch-Gordan coefficients of the group yield a off-diagonal symmetric Majorana submatrix, whose decomposition offers TBM seesaw mixing without fine-tuning the familon vacuum expectation values. All familons in the seesaw sector take crystallographic vacuum alignments for the special case where the familon generating the Majorana submatrix lies along $(1,1,1)$ in vacuum. 

The seesaw sector uses a minimal number of familons. However, the conventional three right-handed neutrino case fails to yield a light neutrino mass spectrum consistent with oscillation data. Introducing a fourth right-handed neutrino, we predict normal ordering of light neutrino masses: $m_{\nu_1} = 27.6$ $\text{meV}$, $m_{\nu_2} = 28.9$ $\text{meV}$ and $m_{\nu_3} = 57.8$ $\text{meV}$. Compared to the sum of neutrino masses restricted by cosmological upper bound ($120$ $\text{meV}$), our prediction yields $114.3$ $\text{meV}$. The model presented in this paper can be falsified with a slight improvement in the cosmological bound.

In Ref.~\cite{rrx2018asymmetric}, we required a phase in the TBM seesaw mixing to reproduce the experimentally determined PMNS angles. In our analysis, this phase arises from the vacuum expectation value of the seesaw familons. Ref.~\cite{CentellesChulia:2019ldn} discusses a different approach where this phase can arise from the residual flavor and generalized $CP$ symmetries \cite{everett2017lepton, *ding2013generalised, *li2016a4, *nishi2013generalized, *sinha2019c, *ding2014lepton, *hagedorn2015lepton, *ding2016generalized, *li2015lepton, *di2015lepton, *ballett2015mixing, *turner2015predictions, *ding2013spontaneous, *feruglio2013lepton,  *feruglio2014realistic, *li2014generalised, *li2015deviation, *lu2017alternative, *king2014lepton, *ding2014generalized, *chen2015neutrino, *ding2014generalised2, *chen2018neutrino, *lu2018quark, *joshipura2018pseudo, *rong2017lepton, *Everett:2015oka, *Lu:2018oxc, *girardi2014generalised, *penedo2017neutrino, *Li:2017abz, *Li:2017zmk, *Lu:2016jit, *Li:2016nap, *Yao:2016zev, *Chen:2016ica, *Chen:2016ptr, *Li:2016ppt, *Ding:2013nsa, *chen2019cp, *chen2019cp2, *barreiros2019combining, *sinha2019phenomenological, *Samanta:2019yeg} of the effective neutrino mass matrix. This phase yields $\cancel{CP}$ phases in the lepton sector, best represented in terms of invariants to avoid ambiguity with many existing definitions. We predict the Jarlskog-Greenberg invariant $|\mc J| = 0.028$ for Dirac $CP$ violation, and Majorana invariants $|\mc I_1| = 0.106$ and $|\mc I_2| = 0.011$. Although no strict bound exists on the Majorana invariants from current experiments \cite{Ge:2016tfx, *Minakata:2014jba}, our prediction for $\mc J$ matches with the current PDG fit, albeit with a sign ambiguity. Light neutrino masses and $\cancel{CP}$ phases make prediction for neutrinoless double-beta decay, with the invariant mass parameter $|m_{\beta \beta }|$ determined to be either $13.02$ or $25.21$ $\text{meV}$ depending on the sign of model parameters. Compared to the latest upper bound ($61$--$165$ $\text{meV}$) from the KamLAND-Zen experiment, both of these are only an order of magnitude away.

We also explore the right-handed neutrino mass spectrum in terms of two parameters. Several curious cases of degeneracy arise for a range of values of the parameters. We think these degeneracies may lead to interesting physics, particularly when one considers the decay of the right-handed neutrinos in the context of leptogenesis. Exploring this is the aim of a future publication.  
\begin{acknowledgments}
M.J.P. would like to thank the Departament de F\'{i}sica T\`{e}orica at the Universitat de Val\`{e}ncia for their hospitality during the preparation of this work. A.J.S. would like to acknowledge partial support from CONACYT project CB-2017-2018/A1-S-39470 (M\'exico). M.H.R., P.R., and B.X. acknowledge partial support from the U.S. Department of Energy under Grant No. DE-SC0010296.
\end{acknowledgments}
\newpage
\appendix

\section{$\mc T_{13}$ Group Theory}
$\mc T_{13} = \mc Z_{13} \rtimes \mc Z_{3}$ has two generators $a$ and $b$, related to the subgroups $\mc Z_{13}$ and $\mc Z_{3}$. These generators are nontrivially related to each other, yielding the presentation
\begin{align*}
    \langle a,b ~|~ a^{13}=b^3=I, bab^{-1}=a^3\rangle.
\end{align*}
Its order is $13 \times 3 = 39$ and it is a subgroup of both $SU(3)$ and $G_2$. 

It has a trivial singlet, a complex singlet (and its conjugate) and two complex triplets (and their conjugates), so that 
\begin{align*}
    1^2 + 1^2 + 1^2 + 3^2 + 3^2 + 3^2 + 3^2 = 39.
\end{align*}
The complex singlet is denoted by $\g{1}'$ and the complex triplets are denoted by $\g{3}_1$ and $\g{3}_2$. 

In this appendix, we list the Kronecker products and Clebsch-Gordan coefficients of $\mc T_{13}$. For further details, see \cite{ishimori2012introduction}.

\subsection{Kronecker Products}
\begin{align*}
\g{1'} \otimes \g{1'} &= \gb{1}', \quad \g{1'} \otimes \gb{1}' = \g{1}\\[1pt]
\g{1'} \otimes \g{3}_i &= \g{3}_i, \quad \gb{1}' \otimes \g{3}_i = \g{3}_i\\
\g{3}_1\otimes\g{3}_1&=\gb{3}_1\oplus\gb{3}_1\oplus\g{3}_2\\[1pt]
\g{3}_2\otimes\g{3}_2&=\gb{3}_2\oplus\gb{3}_1\oplus\gb{3}_2\\[1pt]
\g{3}_1\otimes\gb{3}_1&=\g{1}\oplus\g{1}'\oplus\gb{1}'\oplus\g{3}_2\oplus\gb{3}_2\\[1pt]
\g{3}_2\otimes\gb{3}_2&=\g{1}\oplus\g{1}'\oplus\gb{1}'\oplus\g{3}_1\oplus\gb{3}_1\\[1pt]
\g{3}_1\otimes\g{3}_2&=\gb{3}_2\oplus\g{3}_1\oplus\g{3}_2\\[1pt]
\g{3}_1\otimes\gb{3}_2&=\gb{3}_2\oplus\g{3}_1\oplus\gb{3}_1\\[1pt]
\g{3}_2\otimes\gb{3}_1&=\g{3}_2\oplus\g{3}_1\oplus\gb{3}_1
\end{align*}

\subsection{Clebsch-Gordan Coefficients}
\begin{align*}
\matc{
\ket{1}\\
\ket{2}\\
\ket{3}
}_{\g{3}_1}
\otimes
\matc{
\ket{1'}\\
\ket{2'}\\
\ket{3'}
}_{\g{3}_1} &= 
\matc{
\ket{1} \ket{1'}\\
\ket{2} \ket{2'}\\
\ket{3} \ket{3'}
}_{\g{3}_2} \oplus
\matc{
\ket{2} \ket{3'}\\
\ket{3} \ket{1'}\\
\ket{1} \ket{2'}
}_{\gb{3}_1} \oplus
\matc{
\ket{3} \ket{2'}\\
\ket{1} \ket{3'}\\
\ket{2} \ket{1'}
}_{\gb{3}_1} 
\end{align*}
%%%%%%%%%%%%%%%%%%%%%%%%%%%%%%%%%%%%%%%%%%%%%%%%%%%%%%%%%
\begin{align*}
\matc{
\ket{1}\\
\ket{2}\\
\ket{3}
}_{\g{3}_2}
\otimes
\matc{
\ket{1'}\\
\ket{2'}\\
\ket{3'}
}_{\g{3}_2} &= 
\matc{
\ket{2} \ket{2'}\\
\ket{3} \ket{3'}\\
\ket{1} \ket{1'}
}_{\gb{3}_1} \oplus
\matc{
\ket{2} \ket{3'}\\
\ket{3} \ket{1'}\\
\ket{1} \ket{2'}
}_{\gb{3}_2} \oplus
\matc{
\ket{3} \ket{2'}\\
\ket{1} \ket{3'}\\
\ket{2} \ket{1'}
}_{\gb{3}_2}
%%%%%%%%%%%%%%%%%%%%%%%%%%%%%%%%%%%%%%%%%%%%%%%%%%%%%%%%
\\[4pt]
\matc{
\ket{1}\\
\ket{2}\\
\ket{3}
}_{\g{3}_1}
\otimes
\matc{
\ket{1'}\\
\ket{2'}\\
\ket{3'}
}_{\g{3}_2} &= 
\matc{
\ket{3} \ket{3'}\\
\ket{1} \ket{1'}\\
\ket{2} \ket{2'}
}_{\g{3}_1} \oplus
\matc{
\ket{3} \ket{1'}\\
\ket{1} \ket{2'}\\
\ket{2} \ket{3'}
}_{\gb{3}_2} \oplus
\matc{
\ket{3} \ket{2'}\\
\ket{1} \ket{3'}\\
\ket{2} \ket{1'}
}_{\g{3}_2}
%%%%%%%%%%%%%%%%%%%%%%%%%%%%%%%%%%%%%
\\[4pt]
\matc{
\ket{1}\\
\ket{2}\\
\ket{3}
}_{\g{3}_1}
\otimes
\matc{
\ket{1'}\\
\ket{2'}\\
\ket{3'}
}_{\gb{3}_2} &= 
\matc{
\ket{1} \ket{1'}\\
\ket{2} \ket{2'}\\
\ket{3} \ket{3'}
}_{\gb{3}_1} \oplus
\matc{
\ket{2} \ket{3'}\\
\ket{3} \ket{1'}\\
\ket{1} \ket{2'}
}_{\gb{3}_2} \oplus
\matc{
\ket{2} \ket{1'}\\
\ket{3} \ket{2'}\\
\ket{1} \ket{3'}
}_{\g{3}_1}
%%%%%%%%%%%%%%%%%%%%%%%%%%%%%%%%%%%%%%%%%%%%%%%%%%%%%%%%%
\\[4pt]
\matc{
\ket{1}\\
\ket{2}\\
\ket{3}
}_{\g{3}_2}
\otimes
\matc{
\ket{1'}\\
\ket{2'}\\
\ket{3'}
}_{\gb{3}_1} &= 
\matc{
\ket{1} \ket{1'}\\
\ket{2} \ket{2'}\\
\ket{3} \ket{3'}
}_{\g{3}_1} \oplus
\matc{
\ket{1} \ket{2'}\\
\ket{2} \ket{3'}\\
\ket{3} \ket{1'}
}_{\gb{3}_1} \oplus
\matc{
\ket{3} \ket{2'}\\
\ket{1} \ket{3'}\\
\ket{2} \ket{1'}
}_{\g{3}_2}
%%%%%%%%%%%%%%%%%%%%%%%%%%%%%%%%%%%%%%%
\\[4 pt]
\matc{
\ket{1}\\
\ket{2}\\
\ket{3}
}_{\g{3}_1}
\otimes
\matc{
\ket{1'}\\
\ket{2'}\\
\ket{3'}
}_{\gb{3}_1} &= 
\matc{
\ket{1} \ket{2'}\\
\ket{2} \ket{3'}\\
\ket{3} \ket{1'}
}_{\gb{3}_2} \oplus
\matc{
\ket{2} \ket{1'}\\
\ket{3} \ket{2'}\\
\ket{1} \ket{3'}
}_{\g{3}_2} \\
&\oplus  (\ket{1}\ket{1'}+\ket{2}\ket{2'}+\ket{3}\ket{3'})_\g{1}\\
&\oplus  (\ket{1}\ket{1'}+\omega \ket{2}\ket{2'}+ \omega^2 \ket{3}\ket{3'})_\g{1'}\\
&\oplus (\ket{1}\ket{1'}+\omega^2 \ket{2}\ket{2'}+ \omega \ket{3}\ket{3'})_{\gb{1}'}
%%%%%%%%%%%%%%%%%%%%%%%%%%%%%%%%%%%%%%%%%%%%%%%%%%%%%%%%%%%%%%%%%%%%
\\[4 pt]
\matc{
\ket{1}\\
\ket{2}\\
\ket{3}
}_{\g{3}_2}
\otimes
\matc{
\ket{1'}\\
\ket{2'}\\
\ket{3'}
}_{\gb{3}_2} &= 
\matc{
\ket{2} \ket{3'}\\
\ket{3} \ket{1'}\\
\ket{1} \ket{2'}
}_{\g{3}_1} \oplus
\matc{
\ket{3} \ket{2'}\\
\ket{1} \ket{3'}\\
\ket{2} \ket{1'}
}_{\gb{3}_1} \\
&\oplus  (\ket{1}\ket{1'}+\ket{2}\ket{2'}+\ket{3}\ket{3'})_\g{1}\\
&\oplus  (\ket{1}\ket{1'}+\omega \ket{2}\ket{2'}+ \omega^2 \ket{3}\ket{3'})_\g{1'}\\
&\oplus (\ket{1}\ket{1'}+\omega^2 \ket{2}\ket{2'}+ \omega \ket{3}\ket{3'})_{\gb{1}'}
%%%%%%%%%%%%%%%%%%%%%%%%%%%%%%%%%%%%%%%%%%%%%%%%%%%%%%%%%
\\[4 pt]
\left(\ket{1}\right)_{\g{1}'} \otimes
\matc{
\ket{1'}\\
\ket{2'}\\
\ket{3'}
}_{\g{3}_i} &= 
\matc{
\ket{1}\ket{1'}\\
\omega \ket{1}\ket{2'}\\
\omega^2 \ket{1}\ket{3'}
}_{\g{3}_i}
%%%%%%%%%%%%%%%%%%%%%%%%%%%%%%%%%%%%%%%%%%%%%%%%%%%%%%%%%%%
\\[4 pt]
\left(\ket{1}\right)_{\gb{1}'} \otimes
\matc{
\ket{1'}\\
\ket{2'}\\
\ket{3'}
}_{\g{3}_i} &= 
\matc{
\ket{1}\ket{1'}\\
\omega^2 \ket{1}\ket{2'}\\
\omega \ket{1}\ket{3'}
}_{\g{3}_i}, \qquad \omega^3 = 1.
\end{align*}

\newpage
\section{Alternative Choices for $\varphi_{\mc A}$ and $\varphi_{\mc B}$}\label{App:B}
We chose $\varphi_{\mc A} \sim \gb{3}_2$ and $\varphi_{\mc B} \sim \g{3}_2$ and showed how TBM mixing and normal ordering of light neutrino masses follow from the familon vacuum structure. 

\vspace{0.3cm}
The particular form of $\mc A$ in \eqref{Aforms2} becomes important in Eq.~\eqref{cond1}, which requires $\mc A$ to have the same form as $\mc C \mc P'$. For $\varphi_{\mc A} \sim \gb{3}_1$ and $\varphi_{\mc A} \sim \g{3}_2$, choosing $\mc P' \equiv (2\ 3)$ and $(1\ 3)$, respectively, matches $\mc A$ to $\mc C \mc P'$ and leads to similar results as in Section~\ref{sec:3}.

\vspace{0.3cm}
TBM diagonalization of the seesaw matrix requires the decompositions in Eqs.~\eqref{mydecom} and \eqref{Gdecomp}. Choosing the diagonal form of $\mc B$ in \eqref{Bforms2} implies that in Eq.~\eqref{mydecom} $\mc G$ must be diagonal, which from Eq.~\eqref{Gdecomp} requires $\mc D_b$ to be proportional to $\text{diag}(1,1,1)$. This eventually leads to a completely degenerate light neutrino mass spectrum for the three right-handed neutrino case. Introducing a fourth right-handed neutrino can only correct one of the light neutrino masses, still leaving the other two degenerate, incompatible with oscillation data.

\newpage
\section{The $\mc Z_{n}$ `Shaping' Symmetry} \label{app:shape}

The $SU(5) \times \mc T_{13}$ symmetry allows some operator such as $F\Delta\varphi^{(3)}$ (the detailed list of such operators is too exhaustive) which could perturb the texture. Suppose there is a $\mc Z_n$ symmetry whose purpose is to prohibit these terms.

We use $[~\cdot~]$ to denote the $\mc Z_n$ charges of the respective fields. Our starting point is to define the $\mc Z_n$ charges of the following fields
\begin{equation}\label{shape1}
[F]=a, [T]=b, [H_{\bar 5}]=c, [H_{\overline{\g{45}}}]=d, [\bar N]=e, [\bar{N}_4]=f
\end{equation}

Then the $\mc Z_n$ charges of the rest of fields in the scenario with no $\varphi_v$ can be deduced from the couplings in the Lagrangian in Eq.~\eqref{lagrangian}
\begin{align}
%\begin{aligned}
&[\Delta]=b+c, [\Sigma]=a+d, [\Gamma]=[\Omega]=[\Theta]=b-c, [\Lambda]=a-c, [\varphi^{(t)}_{\g{3}_1}]=c-2b, \nonumber \\
&[\varphi^{(1)}]=[\varphi^{(2)}]=[\varphi^{(4)}]=[\varphi^{(5)}]=-a-b-c, [\varphi^{(6)}]=-a-b-d, [\varphi_{\mc B}]=-2e, \nonumber \\
&[\varphi_z]=-e-f, [\varphi_{\mc A}]=c-a-e,  [\varphi^{(3)}]=[\varphi^{}_{\g{3}_2}]=0.\label{shape2}
%\end{aligned}
\end{align}

It is convenient to focus on the couplings of the familons and define
\begin{align}
\label{a'}&a'=[\varphi_{\mc A}]=c-a-e, \\
&b'=[\varphi^{(t)}_{\g{3}_1}]=c-2b,\\
&c'=[\varphi^{(1)}]=-a-b-c\\
&d'=[\varphi^{(6)}]=-a-b-d\\
&e'=[\varphi_{\mc B}]=-2e\\
\label{f'}&f'=[\varphi_z]=-e-f
\end{align}

To make sure these familons do not mix with each other, they should obey the following constraints:

\begin{align}
a',b',c',d',e',f'\neq 0, \\
2a',2b',2c',2d'\neq 0,  \\
a'\neq \pm b',\pm c',\pm d',\pm e',\pm f', \\
b'\neq \pm c',\pm d',\pm e',\pm f', \\
c'\neq \pm d',\pm e',\pm f', \\
d'\neq \pm e',\pm f',\\
e'\neq -f',\\
e'-2f'=0,\\
d'-b'\neq \pm a', \pm b', \pm c', \pm d', \pm e', \pm f'
\end{align}

\noindent These constraints have no solution for $n < 14$. For $n = 14$ there are many sets of solutions, from which we adopt the following:
$$\{n,a',b',c',d',e',f'\}=\{14,11,1,9,8,4,2\}$$
and using Eqs. (\ref{a'})--(\ref{f'}) we get $$\{a,b,c,d,e,f\}=\{1,1,3,4,5,7\}.$$ Then Eqs. (\ref{shape1}) and (\ref{shape2}) give the $\mc Z_{14}$ charges of the fields in the model.

For the scenario with no $\varphi_z$, we redefine $f'=[\varphi_v]$. In this case there is no solution for $n < 12$. For $n = 12$, there are many solutions, from which we adopt
$$\{n,a',b',c',d',e',f'\}=\{12,11,3,8,10,6,2\}.$$

In either case, there remains an unwanted vertex $\bar\Theta\Omega\varphi^{}_{\g{3}_2}$ allowed for any choice of $n$, which yields the diagram
\begin{equation}
\begin{tikzpicture}[baseline=(a.base)]
\begin{feynman}[small]
\vertex (a);
\vertex [right=of a] (b);
\vertex [right=of b] (c);
\vertex [right=of c] (d);
\vertex [below=of b] (e) {$\varphi^{}_{\g{3}_2}$};
\vertex [below=of c] (f) {$\varphi^{}_{\g{3}_2}$};
\vertex [above left=of a] (i1) {\(T\)};
\vertex [below left=of a] (i2) {\(\varphi^{(t)}_{\g{3}_1}\)};
\vertex [above right=of d] (f1) {\(T\)};
\vertex [below right=of d] (f2) {\(H\)};
\diagram* {
	(i1) -- (a) --[insertion=0.5, edge label=\(\Theta\quad \bar\Theta\)] (b) --[insertion=0.5, edge label=\(\Omega\quad \bar\Omega\)] (c) --[insertion=0.5, edge label=\(\Gamma\quad \bar\Gamma\)] (d) -- (f1),
	(a) -- [scalar] (i2),
	(b) -- [scalar] (e),
	(c) -- [scalar] (f),
	(d) -- [scalar] (f2),
};
\end{feynman}
\end{tikzpicture}
\end{equation}
and contributes $\mc O(\lambda^8)$ terms to the up-type quark mass matrix 
\begin{equation}
Y^{(\frac{2}{3})}=\begin{pmatrix}
2\lambda^8&0&\lambda^8\\
0&\lambda^4&0\\
\lambda^8&0&1
\end{pmatrix}.
\end{equation}
Since it happens at $\mc O(\lambda^8)$, we consider it insignificant.

\newpage
\bibliography{majorana}
\bibliographystyle{apsrev4-1}

\end{document}